\documentclass[twocolumn]{aastex62}

\newcommand{\msun}{{M$_\odot$}}
\newcommand{\Al}{$^{26}$Al}
\newcommand{\Fe}{$^{60}$Fe}
\newcommand{\Ne}{$^{22}$Ne}
\newcommand{\F}{$^{19}$F}
\newcommand{\Cl}{$^{36}$Cl}
\newcommand{\Ca}{$^{41}$Ca}

\received{30 April 2021}
\revised{7 September 2021}
\accepted{8 September 2021 for publication}
\submitjournal{ApJ}

\shorttitle{Nucleosynthesis Yields from Massive Stars}
\shortauthors{Brinkman et al.}

\begin{document}

\title{ALUMINIUM-26 FROM MASSIVE BINARY STARS II. ROTATING SINGLE STARS UP TO CORE-COLLAPSE AND THEIR IMPACT ON THE EARLY SOLAR SYSTEM}

\correspondingauthor{Hannah Brinkman}
\email{hannah.brinkman@csfk.org}

\author{Hannah E. Brinkman}
\affil{Konkoly Observatory, Research Centre for Astronomy and Earth Sciences (CSFK), E\"otv\"os Lor\'and Research Network (ELKH), Konkoly Thege Mikl\'os \'ut 15-17, H-1121 Budapest, Hungary}
\affiliation{Graduate School of Physics, University of Szeged, Dom t\'er 9, Szeged, 6720 Hungary}

\author{J.W. den Hartogh}
\affil{Konkoly Observatory, Research Centre for Astronomy and Earth Sciences (CSFK), E\"otv\"os Lor\'and Research Network (ELKH), Konkoly Thege Mikl\'os \'ut 15-17, H-1121 Budapest, Hungary}

\author{C. L. Doherty}
\affil{Konkoly Observatory, Research Centre for Astronomy and Earth Sciences (CSFK), E\"otv\"os Lor\'and Research Network (ELKH), Konkoly Thege Mikl\'os \'ut 15-17, H-1121 Budapest, Hungary}
\affiliation{School of Physics and Astronomy,
Monash University, VIC 3800, Australia}

\author{M. Pignatari}
\affiliation{E.~A.~Milne Centre for Astrophysics, Department of Physics and Mathematics, University of Hull, HU6 7RX, United Kingdom}
\affil{Konkoly Observatory, Research Centre for Astronomy and Earth Sciences (CSFK), E\"otv\"os Lor\'and Research Network (ELKH), Konkoly Thege Mikl\'os \'ut 15-17, H-1121 Budapest, Hungary}
\affiliation{NuGrid Collaboration, \url{http://nugridstars.org}}
\affiliation{Joint Institute for Nuclear Astrophysics - Center for the Evolution of the Elements}

\author{M. Lugaro}
\affil{Konkoly Observatory, Research Centre for Astronomy and Earth Sciences (CSFK), E\"otv\"os Lor\'and Research Network (ELKH), Konkoly Thege Mikl\'os \'ut 15-17, H-1121 Budapest, Hungary}
\affiliation{ELTE E\"{o}tv\"{o}s Lor\'and University, Institute of Physics, Budapest 1117, P\'azm\'any P\'eter s\'et\'any 1/A, Hungary}
\affiliation{School of Physics and Astronomy,
Monash University, VIC 3800, Australia}

\begin{abstract}
Radioactive nuclei were present in the early Solar System, as inferred from analysis of meteorites. Many are produced in massive stars, either during their lives or their final explosions. 
In the first paper in this series \citep{Brinkman1}, we focused on the production of $^{26}$Al in massive binaries. Here, we focus on the production of another two short-lived radioactive nuclei, $^{36}$Cl and $^{41}$Ca, and the comparison to the early Solar System data. 
We used the MESA stellar evolution code with an extended nuclear network and computed massive (10-80 M$ _{\odot} $), rotating (with initial velocities of 150 and 300 km/s) and non-rotating single stars at solar metallicity (Z=0.014) up to the onset of core collapse. We present the wind yields for the radioactive isotopes $^{26}$Al, $^{36}$Cl, and $^{41}$Ca, and the stable isotopes \F{} and \Ne{}. In relation to the stable isotopes, we find that only the most massive models, $\geq$ 60 \msun{} and $\geq$ 40 \msun{} give positive \F{} and \Ne{} yields, respectively, depending on the initial rotation rate. In relation to the radioactive isotopes, we find that the early Solar System abundances of $^{26}$Al and $^{41}$Ca can be matched with by models with initial masses $\geq$40 \msun{}, while $^{36}$Cl is matched only by our most massive models, $\geq$60 \msun{}. $^{60}$Fe is not significantly produced by any wind model, as required by the observations. Therefore, massive star winds are a favoured candidate for the origin of the very short-lived $^{26}$Al, $^{36}$Cl, and $^{41}$Ca in the early Solar System.
\end{abstract}


\section{Introduction} \label{intro}
Radioactive isotopes with short half-lives of less than a few Myr (hereafter  short-lived radioactive isotopes, SLRs), and specifically the famous case of \Al\, with a half life of 0.72\,Myr \citep[][]{Alhalflife}, but also \Cl{}, \Ca{}, and \Fe{}, with half lives 0.301\,Myr \citep[][]{Clhalflife}, 0.0994\,Myr \citep[][]{Cahalflife}, and 2.62 Myr \citep[][]{FeHalfLife}, respectively, were present in the early Solar System (ESS). Their abundances are  inferred from meteoritic data reporting excesses in their daughter nuclei, for example, the  ESS $^{26}$Al/$^{27}$Al ratio has been measured in calcium-aluminium-rich inclusions (CAIs) to be equal to $(5.23 \pm 0.13) \times 10^{-5}$ \citep{Jacobsen2008}. These radioactive isotopes represent the fingerprint of the local nucleosynthesis that occurred nearby at the time and place of the birth of the Sun. Therefore, they give us clues about the environment and the circumstances of such birth \citep{adams10}.\\
\indent These four isotopes can be made in massive stars and expelled both by their winds, mainly during the Wolf-Rayet (WR) phase of stars with an initial mass $\geq$35 M$_{\odot}$ and/or due to binary interactions \cite{Brinkman1}, and in equal or larger amounts by their final core-collapse supernova (CCSN) (\citet{MeyerClayton2000,Lugaro2018} and Lawson et al. in prep.). While the \Al{}, \Cl{}, and \Ca{} are ejected in significant amounts by both the wind and the CCSN, the amount of $^{60}$Fe in the stellar winds is negligible compared to that in the CCSN ejecta. This is because \Fe{} is produced via neutron captures on the unstable $^{59}$Fe, and for this nucleus to capture a neutron instead of decaying, higher neutron densities ($> 10^{10-11}$ cm$^{-3}$) are required than those produced during core He burning. Therefore, \Fe{} is only produced in carbon shell-burning and in explosive He- and C-burning conditions \citep[][]{LandC2006,tur:10,jones:19} towards the end of the evolution. Instead, large amounts of \Al{} are produced during H burning by proton captures on $^{25}$Mg and expelled by the winds. The majority of the \Al{} is expelled together with $^{35}$Cl and $^{40}$Ca, which are produced during He burning by neutron captures on the proceeding stable isotopes, $^{35}$Cl and $^{40}$Ca, respectively \citep[e.g.,][]{Arnould1997,Arnould2006,Gounelle2012,Brinkman1}.\\
\indent Massive star winds have been suggested as a favoured site of the \Al{} in the ESS  \citep[e.g.,][]{Arnould1997,Arnould2006,Gounelle2012,Gaidos2009,Young2014} also because they do not eject $^{60}$Fe. 
Candidate CCSNe sources of \Al{} predict, instead, a more significant ejection of $^{60}$Fe, leading to $^{60}$Fe/$^{56}$Fe ratios orders of magnitude above the value observed in the ESS of $\simeq 10^{-8}$ \citep{TangDauphas2012,Trappitsch18}. Also, the ESS $^{60}$Fe/$^{26}$Al ratio is roughly two to three orders of magnitude lower than that observed via $\gamma$-ray, which sample the average galactic medium \citep{Diehl2013,Wang2020Iron60}, and predicted by CCSN models \citep{AustinSNe, Sukhbold2016}. This suggests that less $^{60}$Fe was present in the ESS as compared to the galactic average, and/or that extra source(s) of $^{26}$Al were present at the time of the birth of the Sun. \\
\indent \citet{Arnould1997,Arnould2006} considered production of several SLRs, including $^{36}$Cl and $^{41}$Ca, by WR winds in both non-rotating and rotating models and concluded that these could have been the sources of these SLRs in the ESS. More recent studies have focused on $^{26}$Al only \citep{Gounelle2012} and also concluded that WR winds are a possible source. 
In our first paper in this series, \cite{Brinkman1} (hereafter Paper\,I), we also focused on $^{26}$Al 
and investigated how binary interactions between non-rotating massive stars can influence its wind yields. We showed that these interactions can lead to a significant increase in the $^{26}$Al wind yields in stars of masses 10-35\,M$_{\odot}$. For more massive stars, which become WR stars, the effect of binary interaction is almost negligible.\\
\indent In the present work we extend Paper\,I by computing the evolution of stellar models up to core collapse with a larger network of nuclear species and reactions, to calculate the wind yields of the SLRs \Cl{}, \Ca{}, and \Fe{}.  We also include rotation in this study as this impacts stellar evolution and the winds \citep[see, e.g.,][for an overview]{2012RvMP-MM}. Our revision of the production of these SLRs is timely because updates in the ESS values of $^{36}$Cl and $^{41}$Ca have become recently available \citep{Liu2017,Tang2017}. Moreover, the implementation of the mass-loss rates and of rotation represent some of the main uncertainties in the models of massive stars, and the differences obtained with different stellar evolution codes need to be considered carefully. We will therefore compare our results to those available in the literature.\\
\indent With the extended nuclear network and calculations to beyond H burning, we also present new predictions for two stable isotopes that are produced during He burning and can be present in the stellar winds: $^{19}$F and $^{22}$Ne. These are of interest because \cite{MeynetArnouldF192000} have shown that WR stars can contribute significantly to the galactic $^{19}$F abundance while \cite{PalaciosF192005} found that WRs are unlikely to be the source of galactic $^{19}$F, when including updated mass-loss prescriptions and reaction rates. Recently, the discussion around $^{19}$F was rekindled by \cite{Joensson2014b,Joensson2014a,Joensson2017} and \cite{Abia2019}, who re-analysed observations of $^{19}$F and proposed that asymptotic giant branch stars are the most likely source of cosmic $^{19}$F. Still, due to the remaining uncertainty in both the mass-loss prescriptions and the reaction rates \citep[see e.g.][]{Stancliffe2005,Ugalde2008}, WRs cannot be excluded as the sources of galactic $^{19}$F \citep[for a recent overview, see e.g.][]{RydeF192020}.
\indent As for $^{22}$Ne, there are puzzling observations of an anomalous $^{22}$Ne/$^{20}$Ne ratio in cosmic rays, which is a factor of $\sim$5 higher than in the solar wind \citep{Prantzos2012CR}. The comparison to model predictions may be a key to finding the source of cosmic rays in relation to OB associations of massive stars.\\
\indent The paper structure is as follows; In Section\,\ref{method} we describe the method and the physical input of our models. In Section\,\ref{Results1} we discuss the stellar evolution results and all the relevant stellar evolution details for our models, and compare them to results from the literature. In Section\,\ref{Results2} we present the nucleosynthetic yields of our models and compare these to various studies in the literature. In Section\,\ref{sec:ESS} we compare our findings to the abundances of the SLRs in the ESS and discuss which stars are good candidates to explain them. In Section\,\ref{sec:Conclusion} we end with our conclusions.
\section{Method and input physics} 
\label{method}
As in Paper\,I, we have used version 10398 of the MESA stellar evolution code \citep{MESA1,MESA2,MESA3,MESA4} to calculate massive star models with and without the effects of rotation. We have included the extended nuclear network of 209 isotopes within MESA such that the stellar evolution and the detailed nucleosynthesis are solved simultaneously. The input physics we used for the single massive stars is described in the next section. Only the key input parameters and the changes compared to the input physics of Paper\,I are discussed.\\
\indent The inlist files used for the simulations are available on Zenodo under a Creative Commons 4.0 license: \url{https://doi.org/10.5281/zenodo.5497213}.
\subsection{Input physics}
The initial masses of our models are 10, 15, 20, 25, 30, 35, 40, 45, 50, 60, 70, and 80\,M$_{\odot}$. The initial composition used is solar with Z=0.014, following \cite{Asplund2009}. For the initial helium content we have used Y=0.28. Our nuclear network contains all the relevant isotopes for the main burning cycles (H, He, C, Ne, O, and Si) to follow the evolution of the star in detail up to core collapse. All relevant isotopes connected to the production and destruction of $^{26} $Al, $^{36}$Cl, $^{41}$Ca, $^{19}$F, $^{22}$Ne, and $^{60}$Fe are also included into our network. Including the ground and isomeric states of $^{26} $Al, the total nuclear network contains therefore the following 209 isotopes:
n, $ ^{1,2} $H, $ ^{3,4} $He, $ ^{6,7} $Li, $^{7-10}$Be, $ ^{8-11} $B, $ ^{12-14} $C, $ ^{13-16} $N, $ ^{14-19} $O, $ ^{17-20} $F, $ ^{19-23} $Ne, $ ^{21-24} $Na,$ ^{23-27} $ Mg, $ ^{25} $Al, $ ^{26} $Al$ _{g} $,$ ^{26} $Al$ _{m} $,$ ^{27,28} $Al, $ ^{27-33} $Si, $ ^{30-34} $P, $ ^{31-37} $S, $ ^{35-38} $Cl, $^{35-41}$Ar, $^{39-44}$K, $^{39-49}$Ca, $^{43-51}$Sc, $^{43-54}$Ti, $^{47-58}$V, $^{47-58}$Cr, $^{51-59}$Mn, $ ^{51-66} $Fe, $^{55-67}$Co, $^{55-69}$Ni, $^{59-66}$Cu, and $^{59-66}$Zn. Following \citet[and references therein]{Farmer2016} a nuclear network of 204 isotopes is optimal for the full evolution of a star, especially because it includes isotopes that influence $Y_{\rm e}$, which are important for the core collapse \citep[see][]{Heger2000}.\\
\indent We have changed the reaction rate library from NACRE to the JINA reaclib \citep{CyburtJINA2010}, version 2.2. The main difference that will affect the evolution is the $^{14}$N(p,$\gamma$)$^{15}$O rate which is updated to \citep[][]{Imbriani2005}.
For \F{}, we use the \F{}($\alpha$,p)\Ne{} from \cite{Ugalde2008}, included in the JINA reaclib.\\
\indent As in Paper\,I, we have used the Ledoux criterion to establish the location of the convective boundaries. The semi-convection parameter, $ \alpha_{sc} $, was set to 0.1 and the mixing length parameter, $ \alpha_{mlt} $, to 1.5. We make use of overshooting via the ``step-overshoot" scheme with $ \alpha_{ov} $=0.2 for the central burning stages. For better convergence of the models, especially in the later stages of the evolution, we switched off the overshoot on the helium burning shell and the later burning shells and the overshoot on the hydrogen shell was reduced to $ \alpha_{ov} $=0.1.\\
\indent We have also updated our wind mass-loss scheme. For the hot phase (T$_{\rm eff}\geq$ 11kK), we use the prescription given by \cite{Vink2000, Vink2001} and for the cold phase (T$_{\rm eff}\leq$ 1kK) we use \cite{NieuwenhuijzendeJager1990}. For the WR-phase we now use \cite{NugisLamers2000} instead of \cite{Hamann1995}. All phases of the wind have a metallicity dependence $\dot{M}\,\propto\,$Z$^{0.85}$ following \cite{Vink2000} and \cite{VinkdeKoter2005}.\\
\indent We have evolved the stars to the onset of core collapse, using an (iron-)core infall velocity of 300\,km/s as the termination point of our simulations.
\subsection{Rotation}
From observations, we know that massive stars rotate, and often at rates high enough to influence their evolution \citep[see e.g.][for a review]{MandM2000}. In Paper\,I, we did not include rotation because we focused on the impact of binary interactions. Here, we do include rotation but do not consider full binary interactions.
Rotation in MESA is implemented as in \cite{Heger2000}. The two variables f$_{\mathrm{c}}$ and f$_{\mu}$ are set to their commonly used values 1/30 and 0.05 \citep[as calibrated by][]{Heger2000}. We include the Taylor-Spruit dynamo for angular momentum transport, following the implementation of \cite{Heger2005TS}. The Taylor-Spruit dynamo is included because this mechanism allows for efficient transport of angular momentum, which is needed to allow for stellar evolution models to match observed rotation rates in many different stellar objects in different stellar evolutionary phases (see e.g. recent publications of \citealt[][]{Aerts2019} and \citealt[][]{BelczynskiRotation2020}).
We use an initial rotational velocity of 150 and 300 km/s, to cover the rotational velocities observed on the main-sequence which are between 200-250 km/s \citep[][]{Arnould2006}. For the models that include the effects of rotation, the wind will receive a rotational boost. MESA includes the boost as given by \cite{LangerBoost}. However, \cite{MandM2000RotationalBoost} pointed out that some effects have been excluded in this treatment, and therefore we implement the rotational boost as in their Equation 4.30. We follow the implementation for MESA by \cite{ZoltRotationalBoost}, to which we have added the temperature-dependence of the empirical alpha-parameter, determined by \cite{Lamer1995Alpha}\footnote{We implemented a step-function to connect their equation 2 and 3 at log(T$_{\rm eff}$)=4.325, to match the data presented in their Figure 7.} Below log(T$_{\rm eff}$)= 3.90, the alpha-parameter is undefined and we set it to 1, which makes the boost disappear. Above log(T$_{\rm eff}$)=4.7, the alpha-parameter is again undefined. Here we have set it to 0.52, extrapolating the results of \cite{Lamer1995Alpha}.
\subsection{Yield calculations}
Our focus is on the pre-supernova isotopic yields from the winds. To calculate these yields, we integrate over time, because wind mass-loss is a continuous process. For the stable isotopes, there are two yields to consider, the total yield and the net yield. The total yield is calculated as described above. The net yield is the total yield minus the initial abundance present in the star. For the SLRs the net yield is identical to the total yield, because there is no initial abundance present in the stars for these isotopes. The total yields and the relevant initial abundances are presented in Section\,\ref{Results2}.
\section{Stellar Models; results and discussion}
\label{Results1}
\begin{figure}
    \centering
    \includegraphics[width=0.5\textwidth]{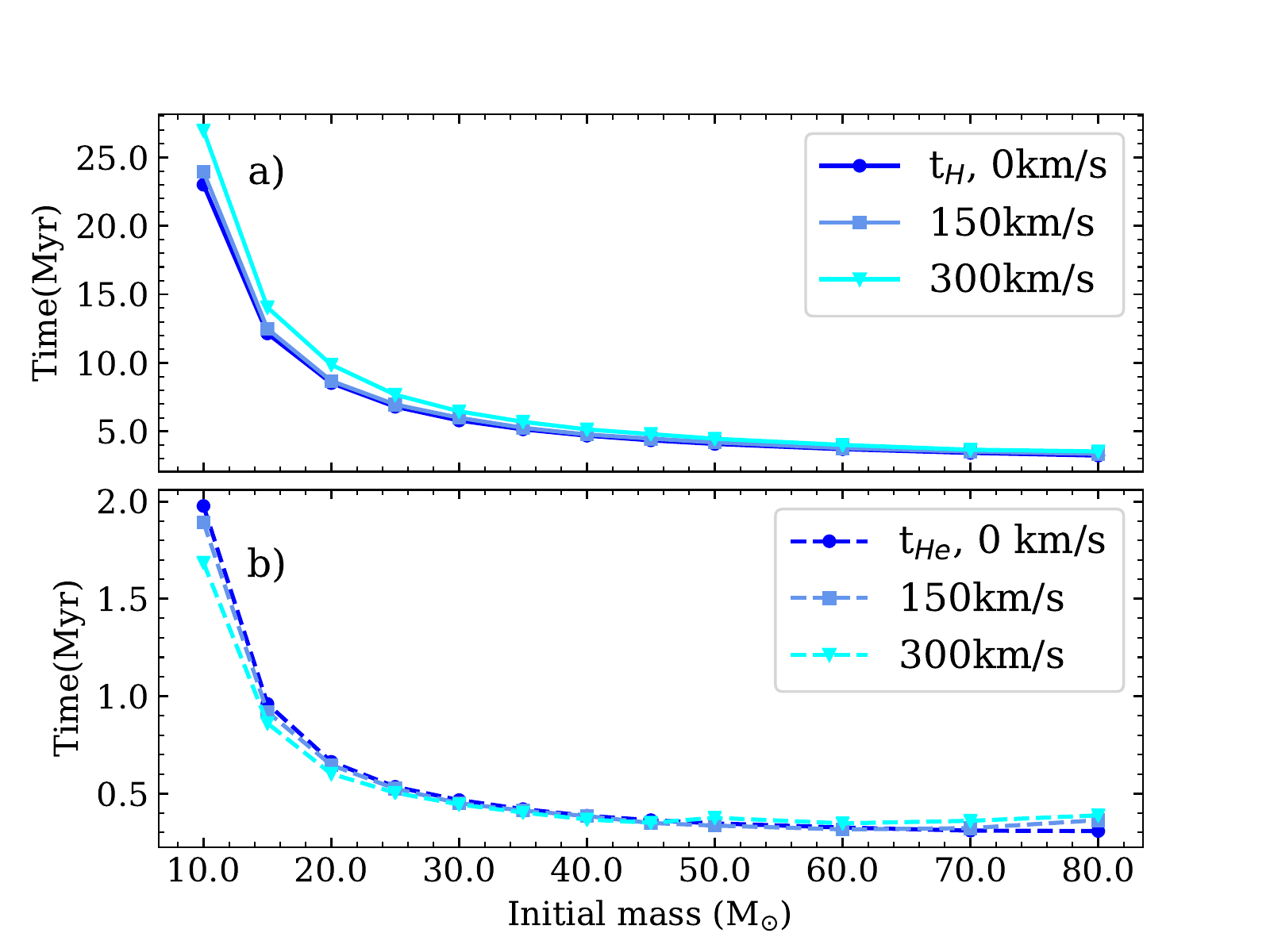}
    \caption{The duration of core hydrogen burning, t$_{\rm H}$ (solid lines in the top panel) and the duration of core helium burning, t$_{\rm He}$ (dashed lines in the bottom panel) for the three rotational velocities as a function of the initial mass.}
    \label{fig:Burning}
\end{figure}

\begin{figure}
    \centering
    \includegraphics[width=0.5\textwidth]{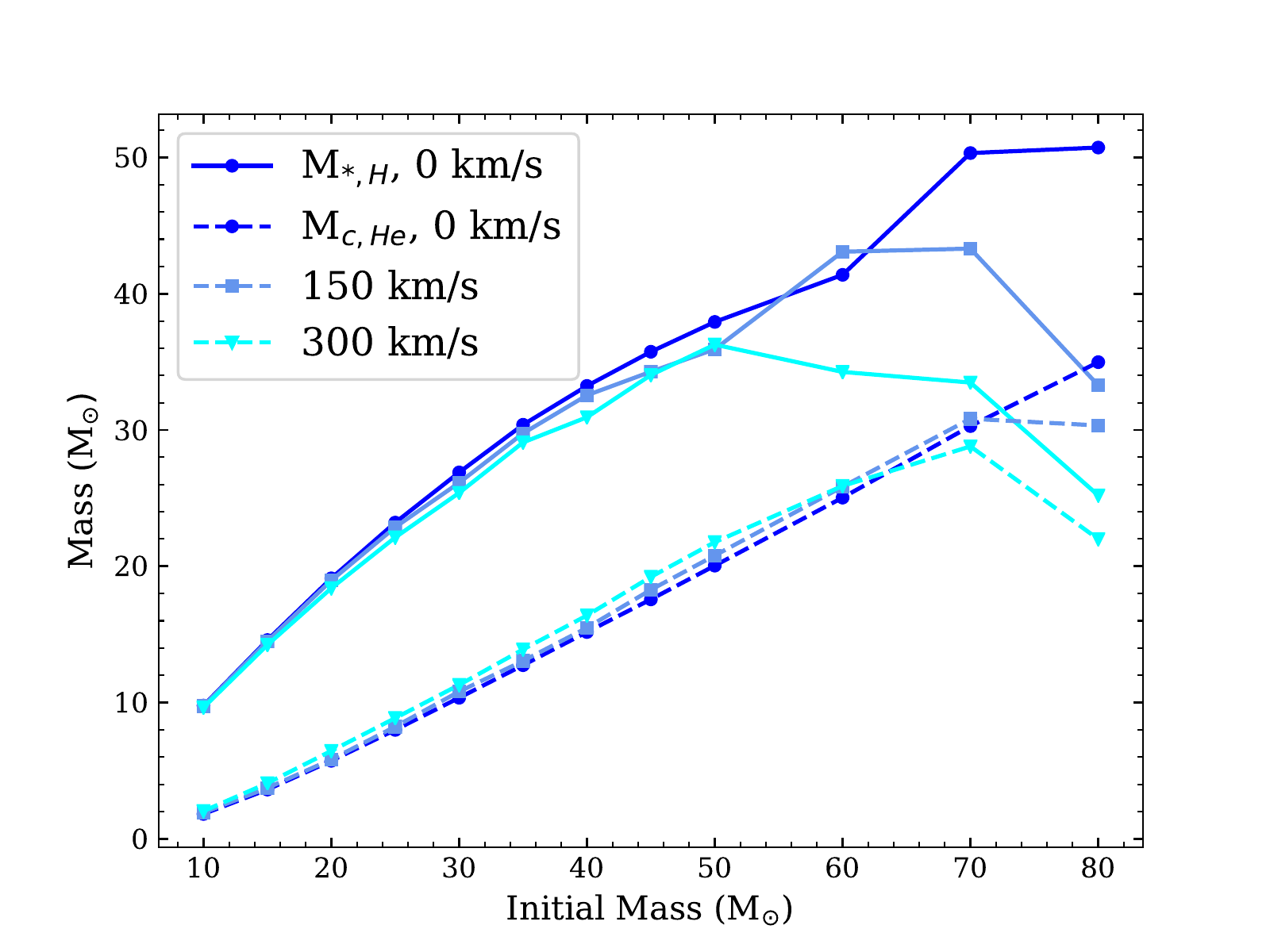}
    \caption{The total stellar mass M$_{*,\rm H}$, (solid lines) and the the hydrogen depleted core mass M$_{*,\rm He}$, (dashed lines) at the end of the main sequence for the three rotational velocities as a function of the initial mass.}
    \label{EndofMS}
\end{figure}

\begin{figure}
    \centering
    \includegraphics[width=0.5\textwidth]{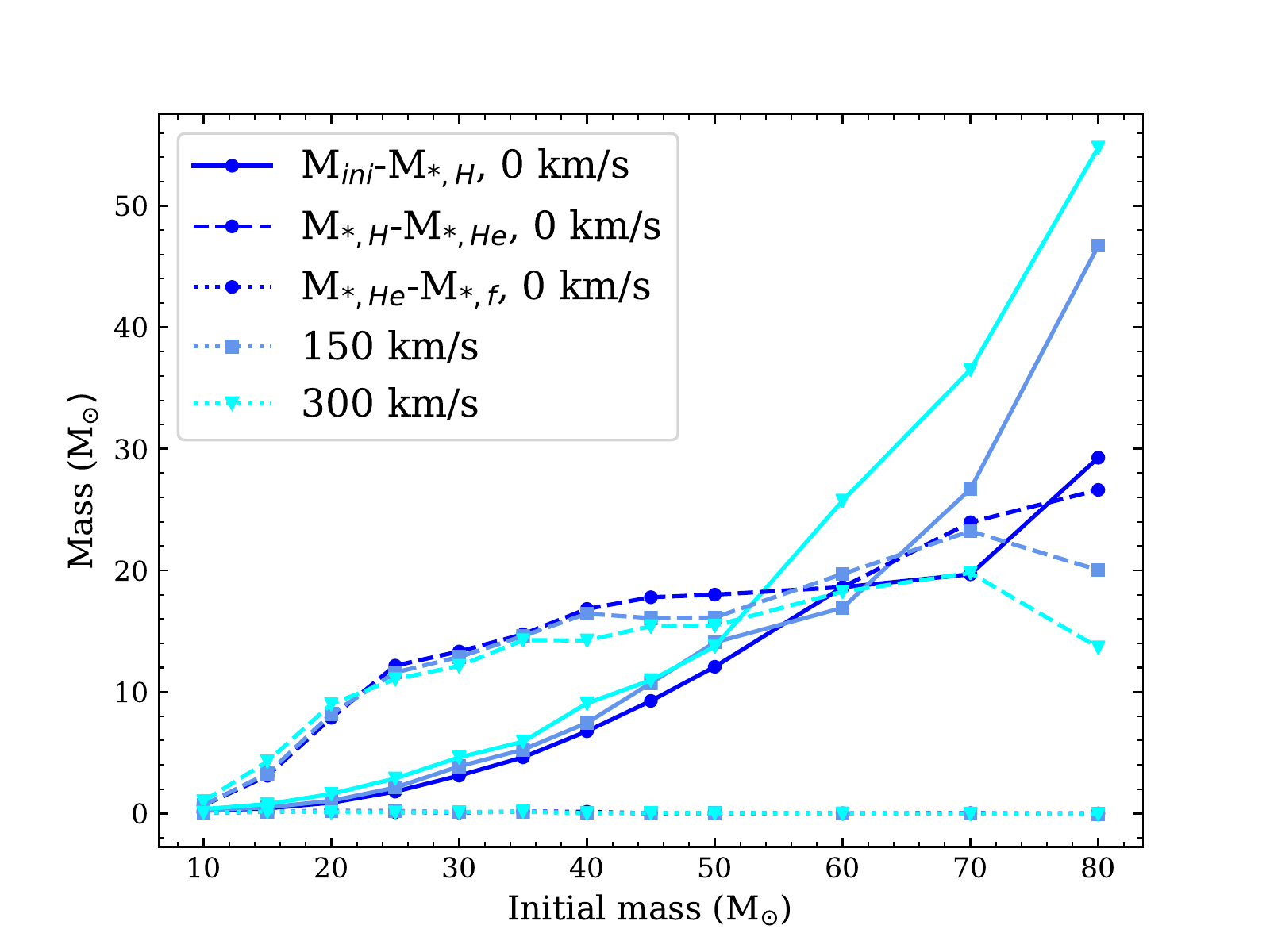}
    \caption{Mass loss for three different phases, the main sequence (solid lines, M$_{\rm ini}$-M$_{*,\rm H}$), helium burning (dashed lines, M$_{*,\rm H}$-M$_{*,\rm He}$) and carbon burning and beyond (dotted lines, M$_{*,\rm He}$-M$_{*,\rm f}$), as a function of the initial mass. The final mass (M$_{*,\rm f}$) is the difference between the initial mass and the total mass loss, M$_{\rm ini}$-$\Delta$M.}
    \label{DeltaM}
\end{figure}

In this section we discuss the stellar evolution details of our models and the impact of rotation on them. In Table\,\ref{StellarInfo}, selected relevant information of our models regarding the stellar evolution is presented: the total stellar mass and core mass at the end of the H, He, and C-core burning, the duration of these burning phases, the total lifetime, the total mass-loss, and the compactness parameter. The start and end points of the different burning phases are defined as in \citet[]{Gotberg2018Condition}\footnote{For completeness: the end of the core H burning stage is defined as when the central helium abundance is larger than 0.98 and the total luminosity produced by nuclear burning is larger than 0.5 the luminosity of the star, the end of He-core burning when the central carbon abundance is 0.4, the end of carbon-burning when the central carbon abundance is 0.01, and the end of oxygen burning when the central oxygen abundance is 0.04.}.
By the end of carbon-burning the mass-loss phase has mostly ended (see Figure \ref{DeltaM}) and all our models have finished this stage. We then continued our models until a core-infall velocity of 300 km/s. In total, 31 out of our 36 models have reached this point, with numerical issues halting the calculations of the remaining 5 models slightly prior to this stage.\\
\indent The duration of hydrogen burning, t$_{\rm H}$ (Column 3 of Table \ref{StellarInfo}), is shown in Figure \ref{fig:Burning}a. 
For all initial masses, t$_{\rm H}$ increases with the rotational velocity. The effect is the strongest at the lower mass end (17$\%$ for 10 \msun{}) and small for the three highest masses of our grid (9$\%$ for 80 \msun{}). The increasing duration of the main sequence is due to rotational mixing. More hydrogen is mixed into the core from the envelope, adding more fuel to the core, and extending this burning phase. Together, this leads to larger hydrogen-depleted cores, M$_{c, \rm He}$, at the end of the main sequence (dotted lines in Figure \ref{EndofMS} and Column 4 in Table \ref{StellarInfo}). The only exceptions to this trend are the two most massive models, 70 and 80 \msun{}, which have a longer main-sequence yet smaller helium-core masses at the end of hydrogen burning when rotating.\\
\indent This is especially noticeable for the highest initial rotational velocity. In these two most massive models, the increased mass loss limits the core growth. With more mass lost from the star, there is less fuel to add to the core and to increase its mass. This is shown by the solid lines in Figures \ref{EndofMS} and \ref{DeltaM}, which give the final mass at the end of the main-sequence, M$_{*,\rm H}$ (Column 5 of Table\,\ref{StellarInfo}) and the mass loss on the main sequence (M$_{\rm ini}$-M$_{*,\rm H}$), respectively. For the models below $\sim$50 \msun{}, there is little difference between in the mass losses between the rotating and non-rotating models. The difference in final mass at the end of the main-sequences ranges between 1-4$\%$ for these stars. Above $\sim$50 \msun{}, the extra mass loss for the rotating models becomes more significant, up to 50$\%$ more for our most massive model.
\\
\indent The mass-loss between the end of the main-sequence and the end of helium burning, M$_{*,\rm H}$-M$_{*,\rm He}$, represented by the dashed lines in Figure \ref{DeltaM}, is comparable between the rotating and non-rotating models with the same initial mass. The exceptions are again the two most massive models, for which the mass loss between the end of H- and the end of He-burning reduces with the increasing rotational velocity. This is because these models have already lost more mass on their main-sequence (solid lines in Figure \ref{DeltaM}), and as a consequence experience less mass loss afterwards.\\
\indent As a result of the larger core masses at the end of hydrogen burning, the helium burning life-time, t$_{\rm He}$, becomes shorter for most of the models with rotation (Figure \ref{fig:Burning}b and Table \ref{StellarInfo}). This is because the heavier the cores, the faster the burning. The three most massive models, however, the rotating models have smaller helium core masses at the end of hydrogen burning compared to their non-rotating counterparts. For these masses, the rotating models have longer t$_{\rm He}$ than their non-rotating counterparts. The turn-over point is around $\sim$50 \msun{}, as can be seen in Figure \ref{fig:Burning}b.\\
\indent After helium burning barely any mass is lost from the stars (dotted lines in Figure \ref{DeltaM}). The helium-depleted core at the end of helium burning, M$_{c,\rm C}$ (Column 7 in Table \ref{StellarInfo}) increase in mass for the models below $\sim$50 \msun{}. For three highest masses however the mass of the helium-depleted cores decreases with the initial rotational velocity, which is a direct result of the smaller helium cores earlier.\\
We note that the mass of the carbon depleted core, M$_{c,\rm O}$ (reported in Column 10) is very sensitive to mixing in the final phases of the stellar evolution and a small fluctuation in the mixing can easily alter this value.\\
\indent The mass loss during the main sequence always increases with higher rotational velocities (solid lines in Figure \ref{DeltaM}). The same applies for the total lifetime of the star, t$_{\rm tot}$ (Column 9 of Table \ref{StellarInfo}). On the other hand, the total mass loss during the whole evolution, $\Delta$M, (see Table \ref{StellarInfo}) is not always larger for the higher rotation rates.\\
\indent The final column of Table \ref{StellarInfo} gives the compactness-parameter, $\xi_{2.5}$. This parameter, as defined by \cite{CompactnessOconnor} in their Equation 10, determines how compact the core of the star is just before the collapse and therefore how difficult it is to explode the star. The compactness is sensitive to small changes in the structure, and is therefore strongly dependent on the model parameters and the codes used (see e.g., \citealt{Sukhbold2016,LandC2018} and \citealt{Schneider2021Compactness}).
\subsection{Comparison to other data sets}
We compare our results primarily to those of \citealt{Ekstrom2012}, calculated using the GENEC stellar evolution code (hereafter E12), and \citealt{LandC2018} calculated with the FRANEC stellar evolution code (hereafter LC18), for three reasons. The first is that both studies present rotating and non-rotating massive star models until the late phases of evolution. The second is that the implementation of rotation differs in these codes compared to what is used in MESA. Specifically, the treatment of rotation in MESA is based on a diffusive approximation, while in GENEC and FRANEC it is based on diffusion–advection approach \citep{1998maeder,MESA2}, with it well established that the impact of rotation on stellar model computations varies depending on which rotational mixing approach is used \citep[see, e.g,][]{MandM2000}. Finally, all the stellar evolution details from these two studies are publicly available, though for E12 we do not have yields for \Cl{}, \Ca{}, and \Fe{}. We compare our models to those from the literature at solar metallicity (taken to be either Z=0.014, or Z=0.02, depending on the source)\\
\indent The E12 models have an initial rotational velocity of 0.4$\omega_{\rm crit}$ (corresponding to $\sim$260-350 km/s). The LC18 models have the same initial rotational velocities of 150 and 300 km/s, as our models.\\
\indent Our models show similar trends as in those works, such as the extended duration of the main-sequence and  the increased mass-loss in the early phases of the evolution. Overall, all the different models show similar main-sequence lifetimes. However, the choice of mass loss prescription makes a big difference between the sets. The LC18 models lose much more mass in the early phases when rotation is included. When looking at their Figure 7, their lowest mass models including rotation move up almost vertically in the Hertzsprung-Russell diagram at the end of the main-sequence. This leads to a strong increase in the mass-loss, which is not seen in our models nor in those by E12.
\begin{table*}[t]
\caption{Selected details of the evolution of our stellar models. M$_{\rm ini}$ is the initial mass in M$_{\odot}$. V$_{\rm ini}$ is the initial rotational velocity in km/s. t$_{\rm H}$, t$_{\rm He}$, and t$_{\rm tot}$ are the duration of hydrogen burning, helium burning, and the total evolution time in Myr, respectively. M$_{*,\rm H}$, M$_{*,\rm He}$, and M$_{*,\rm C}$ are the masses of the stars at the end of their respective burning phases. M$_{c,\rm He}$, M$_{c,\rm C}$, and M$_{c, \rm O}$ are the masses of the hydrogen-depleted core, the helium-depleted core, and the carbon depleted core at the end of the corresponding burning phases in M$_{\odot}$. $\Delta$M is the total mass lost in  M$_{\odot}$. $\xi_{2.5}$ is the compactness of the star at the final model.}
\begin{center}\begin{tabular}{cc|ccc|ccc|ccc|cc}
\hline 
M$ _{\rm ini} $ &  V$ _{\rm ini} $& t$ _{\rm H} $ & M$ _{c, \rm He} $ &M$_{*,\rm H}$&t$ _{\rm He} $ & M$ _{c,\rm C} $&M$_{*,\rm He}$ &t$ _{\rm tot} $ & M$ _{c,\rm O} $& M$_{*,\rm C}$& $  \Delta$M& $\xi_{2.5}$\\
(M$ _{\odot}$) & (km/s) & (Myr)& (M$ _{\odot} $)& (M$ _{\odot} $)& (Myr)& (M$ _{\odot} $)& (M$ _{\odot} $)&(Myr) & (M$ _{\odot} $) &(M$ _{\odot} $)& (M$ _{\odot} $)& -\\
\hline
10 & 0$^{1}$ &23.00 &1.83 & 9.78 &1.98 &1.52 &9.11 &25.37 &1.38 &9.04 &0.96 &6.09e-3\\
 & 150$^{1}$ &23.95 &1.92 &9.75 &1.89 &1.58 &9.06 &26.21 &1.34 &8.99 &1.01 &7.69e-3\\
 & 300 & 26.96 &2.04 &9.64 &1.69 &1.84 &8.62 &28.99 &1.91 &8.55 &1.45 &0.017\\
\hline
15   & 0 & 12.15 &3.62 & 14.58 &0.96 &3.41 &11.49 &13.26 &1.71 &11.35 &3.65 &0.098 \\
   & 150 &12.49 &3.73 & 14.50 &0.92 &3.54 &11.22 &13.56 &1.60& 11.08 &3.91 &0.078\\
   & 300 &14.05 &4.09 &14.23 &0.86 &4.04 &9.96 &15.03 &1.66 &9.80 &5.20 &0.086\\
\hline 
20   & 0 &8.53 &5.73&19.12 &0.66 &5.66 &11.24 &9.29 &2.49& 11.02 &8.98 &0.23\\
   & 150 &8.68 &5.82 &18.96 &0.65 &5.83 &10.77 &9.43 &1.92&10.55 &9.44 &0.16\\
   & 300$^{1}$ &9.87 &6.45 &18.39 &0.60 &6.62 &9.39 &10.55 &2.99 &9.28 &10.71 &0.40\\
\hline
25   & 0$^{1}$ &  6.80 &7.99& 23.20 &0.53 &8.13 &11.04 &7.41 &1.71 & 10.92 &14.07 &0.12\\
   & 150 & 6.96 &8.23 &22.87 &0.53 &8.06 &11.27 &7.56 &1.77&11.04 &13.95 &0.11\\
   & 300 & 7.68 &8.86&22.12 &0.50 &8.63 &11.06 &8.25 &1.90 &10.96 &14.03 &0.11\\
\hline
30   & 0$^{1}$ &5.80 &10.35&26.89 &0.47 &10.73 &13.56 &6.32 &2.44 &13.47 &16.52 &0.23\\
   & 150 & 6.00 &10.80 &26.13 &0.45 &10.70&13.25 &6.50 &2.23 &13.12 &16.87 &0.22\\
   & 300 & 6.47 &11.29 &25.39 &0.44 &10.81 &13.24 &6.96 &2.16 &13.12 &16.86 &0.19\\
\hline 
35   & 0 & 5.15 &12.74 &30.38 &0.42 &12.95 &15.64 &5.62 &3.00 &15.46 &19.51 &0.18\\
   & 150 &5.27 &13.05&29.75 &0.41 &12.55 &15.15 &5.73 &2.79&15.00 &19.98 &0.23\\
   & 300 &5.70 &13.90 &29.09 &0.40 &12.12 &14.83 &6.15 &2.69& 14.65 &20.33 &0.29\\
\hline 
40   & 0 & 4.69 &15.17 &33.23 &0.39 &13.63 & 16.40 &5.12 &3.21 &16.24 &23.74 &0.19\\
   & 150 & 4.78 &15.49&32.55 &0.38 &13.21&16.11 &5.20 &3.07& 16.04 &23.93 &0.23\\
   & 300 & 5.15 &16.39 &30.94 &0.37 &13.72 &16.70 &5.56 &3.09 &16.65 &23.32 &0.26\\
\hline
45   & 0 &4.35 &17.57 & 35.74 &0.36 &14.85 &17.95 &4.76 &3.43 &17.90 &27.06 &0.26\\
   & 150 & 4.48 &18.28&34.28 &0.35 &15.26 & 18.20 &4.88 &3.28 & 18.13 &26.82 &0.26\\
   & 300 & 4.81 &19.22 &34.03 &0.35 &15.57 &18.63 &5.20 &3.62 &18.55 &26.41 &0.31\\
\hline
50   & 0 &4.09 &20.05 &37.93 &0.35 &16.84 &19.92 &4.48 &3.77& 19.87 &30.07 &0.31\\
   & 150 & 4.21&20.79 & 35.93 &0.34 &16.68 & 19.80 &4.59 &3.53 & 19.71 &30.23 &0.30\\
   & 300 & 4.47 &21.77 &36.26 &0.38 &17.70 & 20.81 &4.85 &3.99& 20.73 &29.21 &0.33\\
\hline 
60   & 0 &3.71 &25.04 &41.39 &0.33 &19.49 &22.76 &4.07 &4.53 & 22.67 &37.25 &0.31\\
   & 150 &3.78 &25.84 &43.08 &0.32 &20.09 &23.37 &4.14 &4.60 &23.29 &36.63 &0.32\\
   & 300 & 4.02 &25.89 &34.26 &0.35 &13.19 &16.01 &4.40 &3.15 &15.93 &43.99 &0.23\\
\hline
70   & 0 & 3.43 &30.29 &50.32 &0.31 &22.86 &26.36 &3.78 &15.14$^{2}$ &26.25 &43.65 &0.46\\
   & 150 & 3.53& 30.82&43.31 & 0.32& 16.89 &20.08 &3.89 &3.88 &19.99 &49.92 &0.32\\
   & 300 & 3.67 &28.80 &33.48 &0.36 &11.10 &13.68  &4.07 &2.74&13.60 &56.30 &0.29\\
\hline
80 &0 &3.24 &34.98 &50.72 &0.31 &20.68 &24.10 &3.58 &4.84 &23.99 &55.90 &0.36\\
   & 150 &3.36 &30.33 &33.28 &0.36 &10.67&13.24 &3.76 &2.45 & 13.17 &66.72 &0.25\\
   & 300 &3.54 &21.99 &25.20 &0.39 &9.10 &11.54 &3.97 &2.06 &11.46 &68.42 &0.16\\
\hline
\end{tabular}\end{center}
$^{1}$ This run was terminated before core collapse was reached due to numerical difficulties, however, negligible mass loss is expected after this point.
$^{2}$ This run experienced computational difficulties in the final phases, leading to a much larger M$_{c,\rm O}$ than for any of the other models.
\label{StellarInfo}
\end{table*}

\section{Stellar Yields; results and discussion}
\label{Results2}
In this section we present and discuss the wind yields for \Al{}, \F{}, \Ne{}, \Cl{}, and \Ca{} from our models and compare them to other studies in the literature. We do not present the \Fe{} yields in quantitative detail because our models confirm all the previous results that this isotope is not ejected in the winds (the maximum yield we found is 4.55$\times$10$^{-10}$ \msun{}for the non-rotating 70 \msun{} model). In Table \ref{Yields} the yields of the five isotopes and the initial abundances of $^{19}$F and $^{22}$Ne are presented. The complete set of wind yields for all isotopes and models presented here, are available on Zenodo under a Creative Commons 4.0432 license: \url{https://doi.org/10.5281/zenodo.5497258}.
\indent As discussed in Section \ref{Results1}, for the low mass end, 10-35 \msun{}, of the stars we investigate, most of the mass is lost between the main-sequence and the onset of helium-burning. At the high mass end, $ \sim$35-80 \msun{}, instead, the stars become WR stars and continue to lose mass even during and shortly after helium burning, stripping away not only the hydrogen-rich envelope, but also the top of the hydrogen-depleted core. This strongly impacts the yields of these stars, especially for the isotopes synthesized after hydrogen burning.\\
\subsection{Short lived radioactive isotopes}
\label{slrs}
\begin{figure*}
    \centering
    \includegraphics[width=\linewidth]{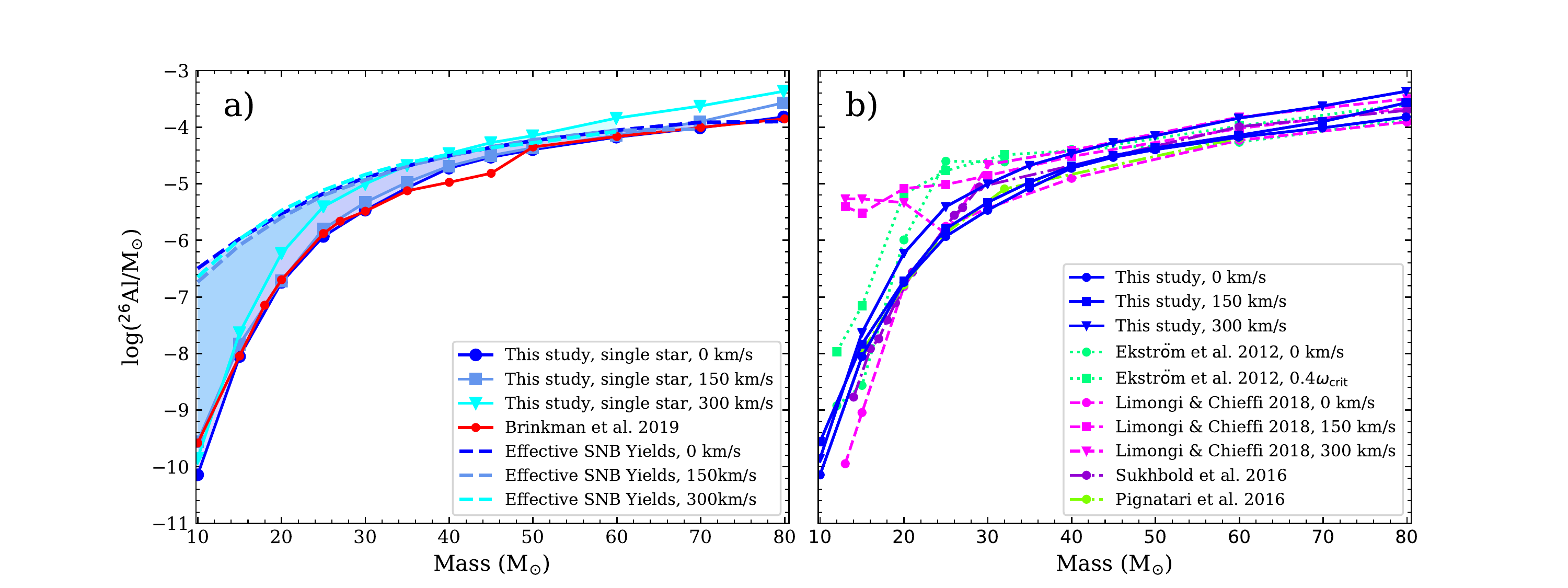}\\
    \includegraphics[width=\linewidth]{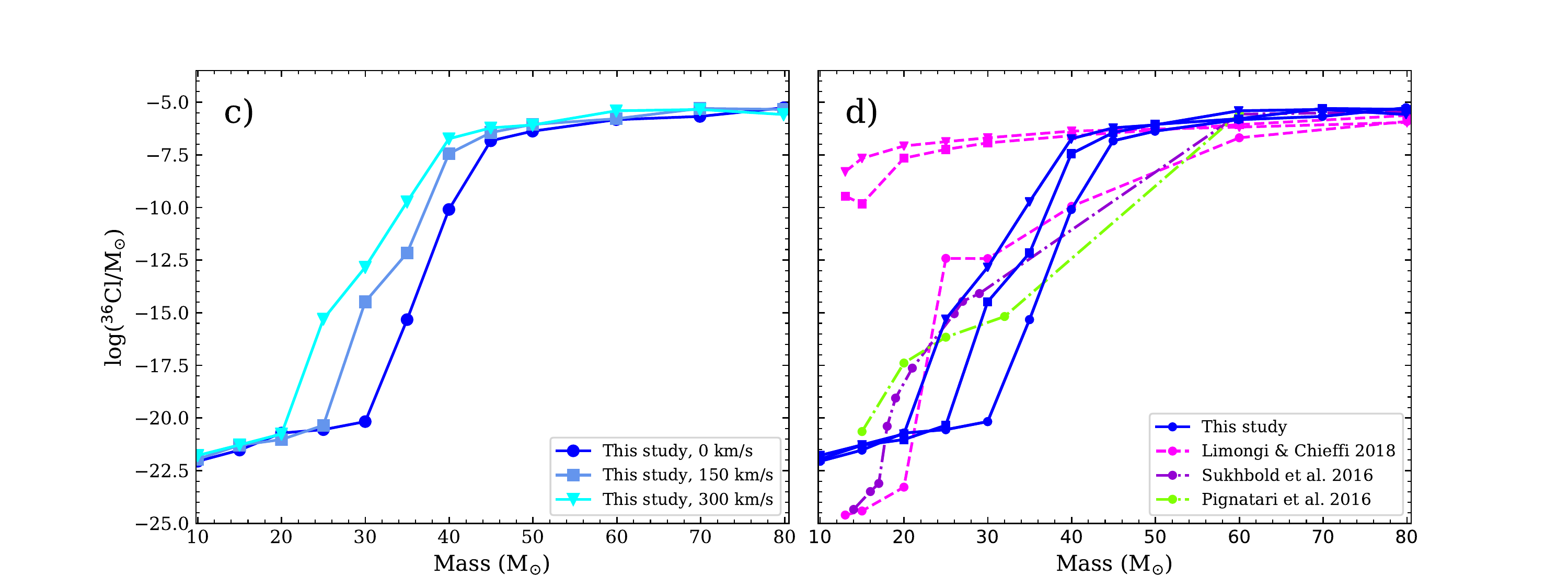}\\
    \includegraphics[width=\linewidth]{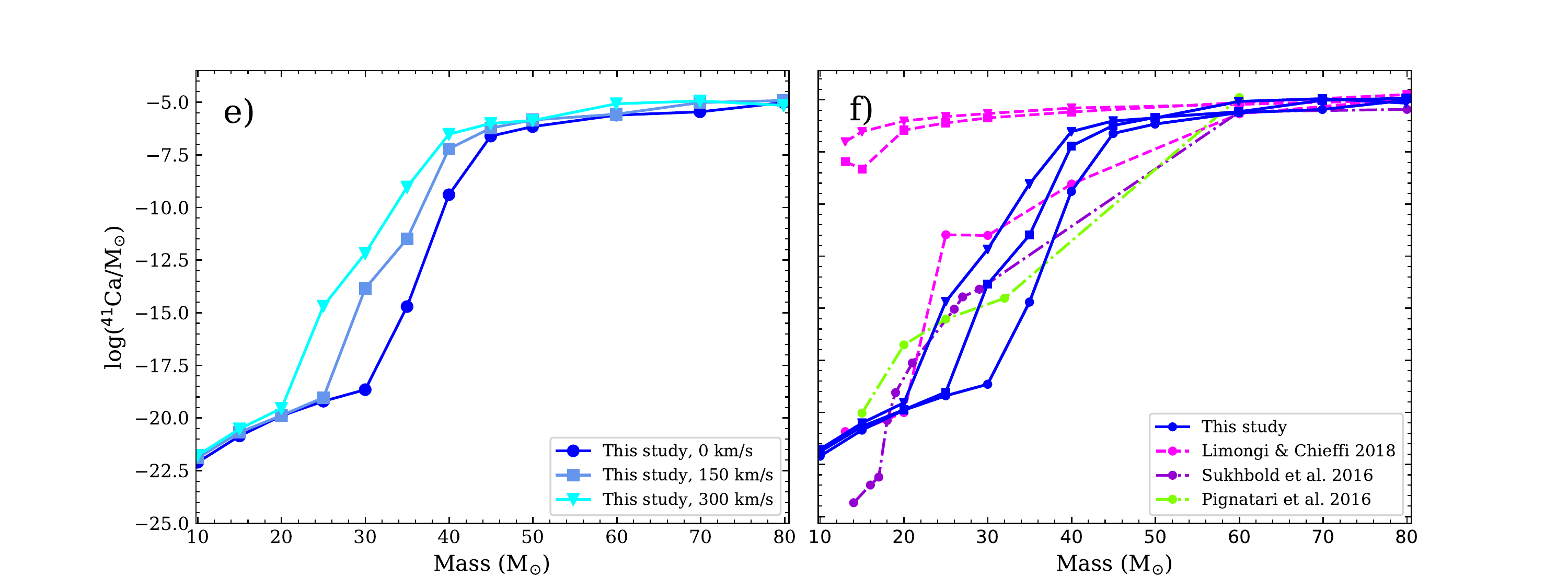}
    \caption{The left panels show our SLR yields for the single star models (solid lines with symbols) for the three different initial rotational velocities. In panel (a), \Al{}, the red line shows the yields from Paper\,I. The dashed lines give the "effective binary yields'' defined as SNB yields averaged on the period range given in Paper I assuming a flat period distribution. These effective binary yields are presented to help visualise the effect of binary interactions. The shaded area between the single star models and the effective binary yields therefore represents the results from the potential parameter space covered by binary systems. The right panels show the single star models of this study together with the wind yields of various other studies, as listed in the legend.}
    \label{fig:SLRs}
\end{figure*}
\begin{figure*}
    \centering
    \includegraphics[width=0.49\linewidth]{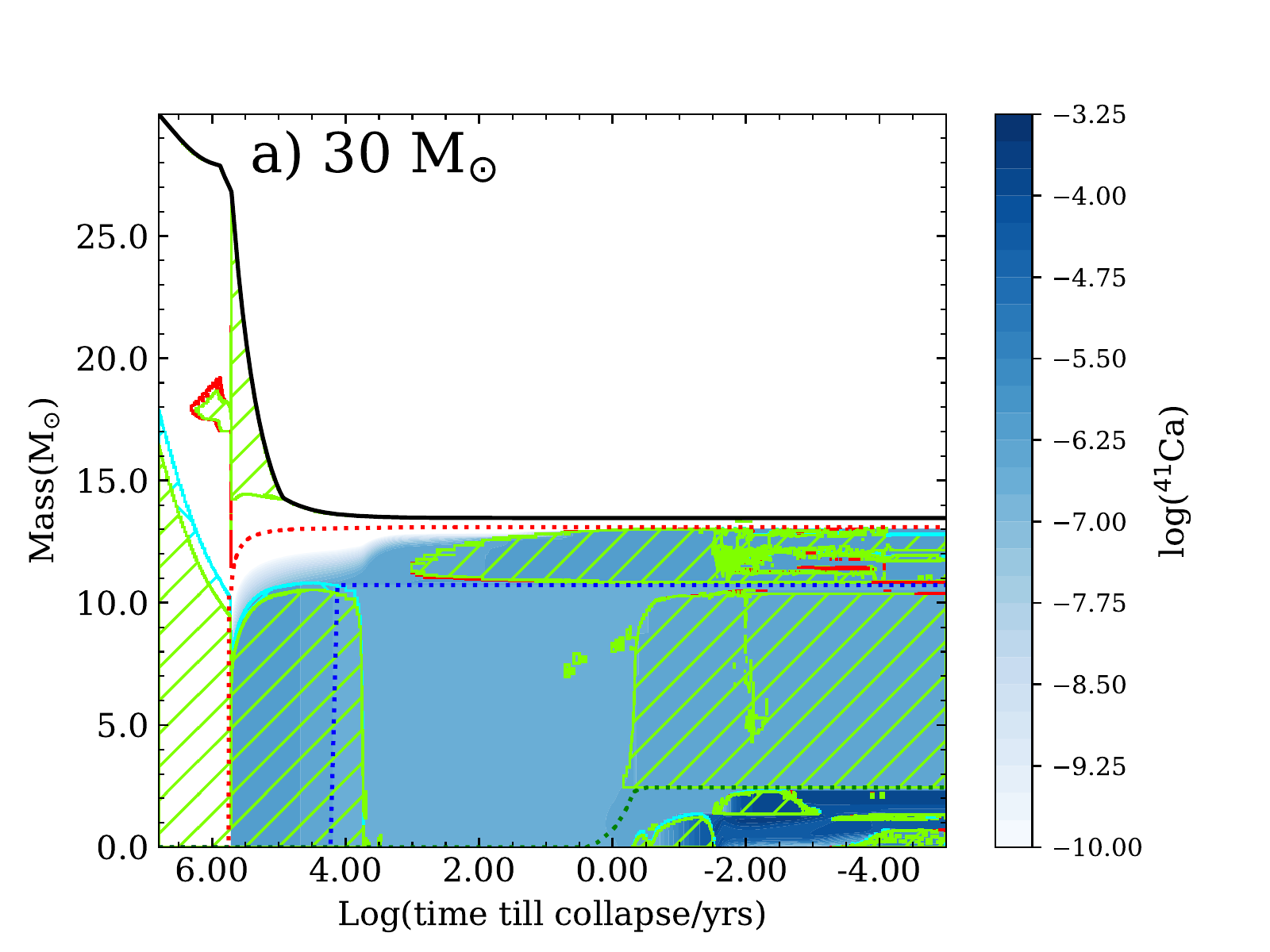}
    \includegraphics[width=0.49\linewidth]{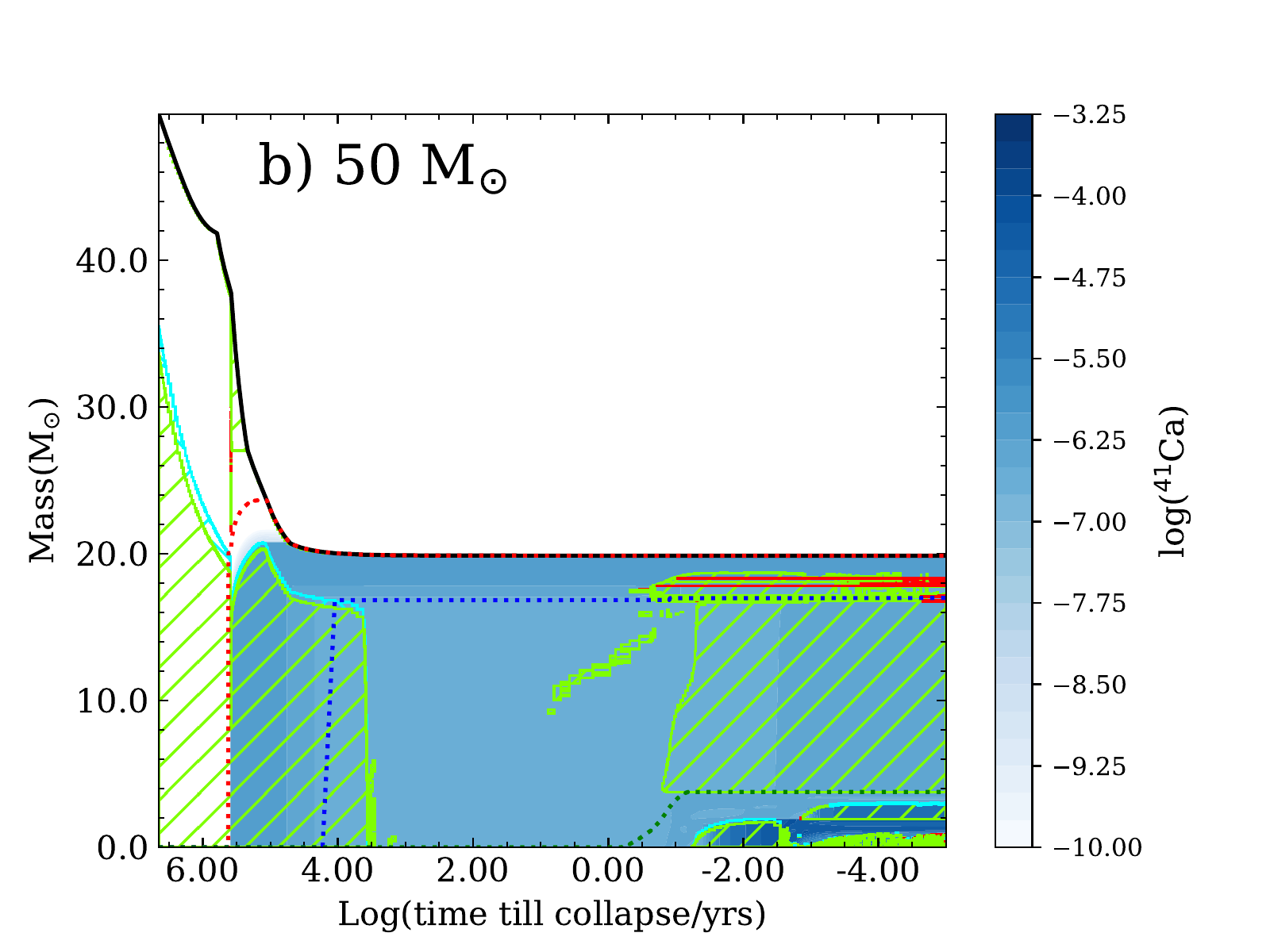}
    \caption{Kippenhahn diagrams for a 30 \msun{} star in the left panel and for a 50 \msun{} the right panel, both non-rotating. The horizontal axis shows the time left until core collapse. The colour-scale shows the \Ca{} mass-fraction. The green-hatched areas represent convective regions, the blue hatched areas, overshoot, and the red areas semi-convection. The red dotted line is the size of the hydrogen-depleted core.\newline}
    \label{fig:CaKHDs}
\end{figure*}

\indent For \Al{} (Figure \ref{fig:SLRs}a), the non-rotating yields are comparable to those from Paper\,I, with the exceptions of the 10, 40, and 45 \msun{} models. The reason is that the 10 \msun{} model loses more mass in our earlier study due to a longer main sequence lifetime, leading to a slightly higher yield. The 40 and 45 \msun{} models are evolved until core-collapse here, while earlier they did not finish helium burning. This leads to a larger mass-loss and therefore a higher yield.
The largest impact of the SNBs is at the lower mass-end, 10-35 M$_{\odot}$ (as shown in Paper I). However here we also find that, because both rotation and binary interaction lead to an increased mass-loss at similar points in the evolution, the impact of the SNBs becomes relatively smaller with increasing initial rotational velocity.
The effect of binary interactions becomes negligible for models with initial masses of 40-45 \msun{}, depending on the initial rotational velocity. Overall, our \Al{} yields are only mildly sensitive to rotation and the wind prescription, since we changed the latter from \citet{Hamann1995} to \citet{NugisLamers2000} for the WR phase. The effect of the binary interactions is stronger than the effect of rotation or changing the wind.\\
\indent Figures\,\ref{fig:SLRs}c and \ref{fig:SLRs}e show the $^{36}$Cl and \Ca{} wind yields for our models. Unlike for \Al{}, where the yield increases gradually with increasing mass, these yields show a sharp rise of almost 15 orders of magnitude at masses between 20-30 \msun{}, depending on initial rotational velocity. This is easily understood when considering Figure\,\ref{fig:CaKHDs}, which shows the KHDs for non-rotating 30 \msun{} and 50 \msun{} models, with the \Ca mass-fraction on the colourscale. Prior to core helium burning, there is no \Ca{} present within these stars. The abundance of \Ca{} (and \Cl{}) increases due to neutron captures during He burning on $^{40}$Ca (and $^{35}$Cl), and it is also destroyed by neutron capture reactions \citep[see, e.g.,][Figure 7]{Lugaro2018}. For the 30 \msun{} model, this \Ca{} (and \Cl{}) barely reaches the surface, leading to a very low yield. However, for the 50\,\msun{} model the top layers of the helium burning core are stripped away, leading to a more significant \Ca{} (\Cl{}) yield.  For the rotating models, the WR-phase is reached at a lower initial mass, and the increase in the yields also moves towards these lower initial masses. This means that these yields are highly sensitive to the wind prescription used, especially for the WR phase.
\subsection{Stable isotopes}
\begin{figure*}
    \centering
    \includegraphics[width=\linewidth]{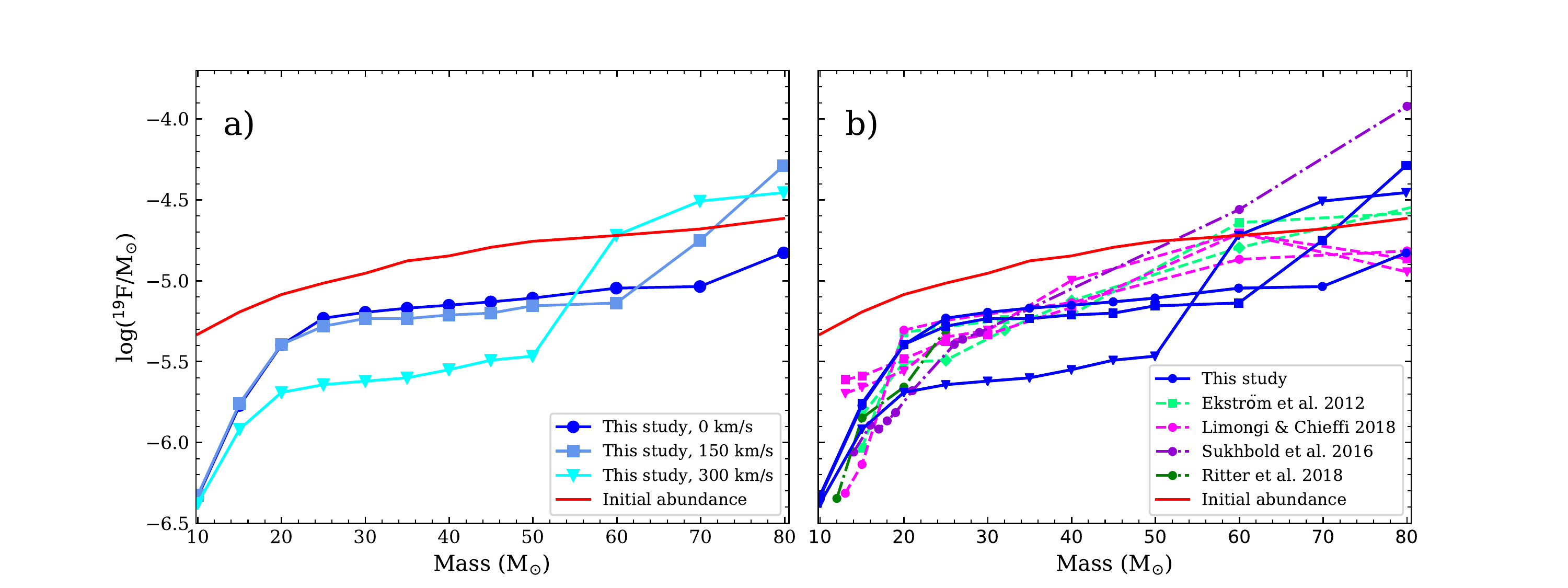}\\
    \includegraphics[width=\linewidth]{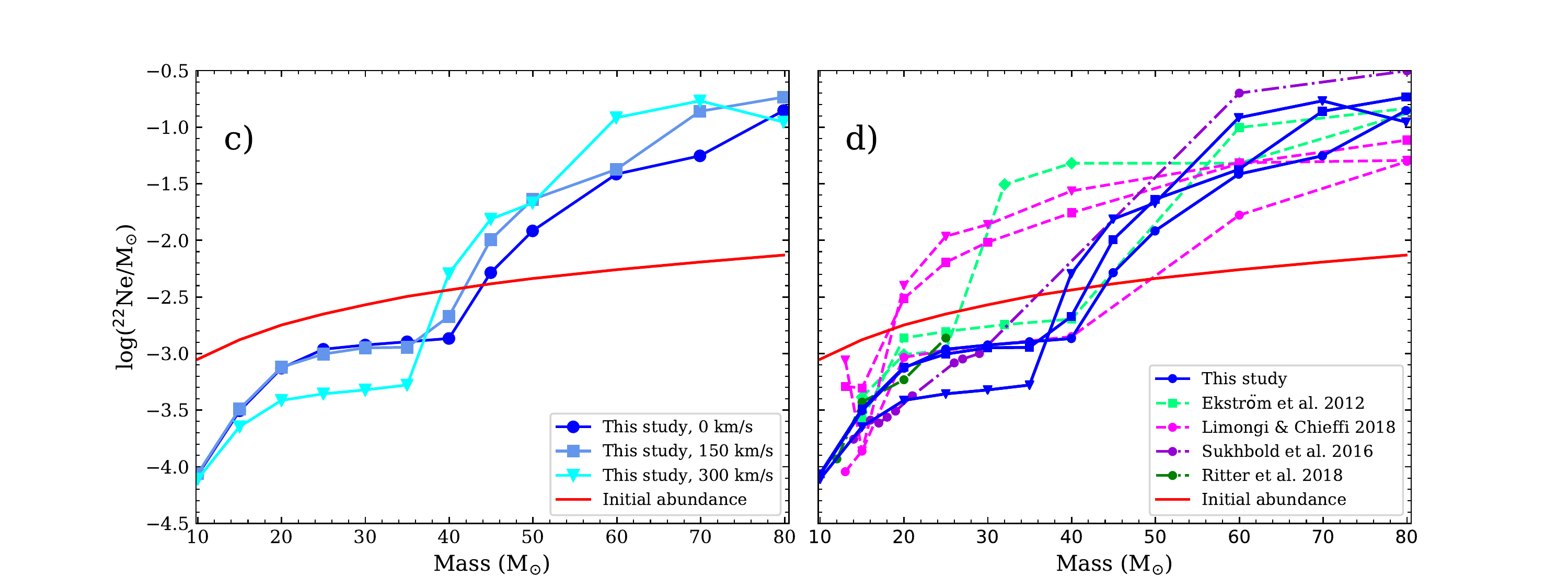}
    \caption{The left panels show our $^{19}$F (top) and \Ne{} (bottom) yields and the initial abundances. The red line indicates the initial abundance of the models. The right panels show the stellar models of this study together with the yields of various other studies.}
    \label{fig:Stables}
\end{figure*}

Unlike the SLRs, the stable isotopes \F{} and \Ne{} are already present in the star at the time of its birth. Both these isotopes are typically destroyed by proton captures during hydrostatic H-burning as part of the CNO and NeNa-cycles through the \F{}(p,$\alpha$)$^{16}$O and \Ne{}(p,$\gamma$)$^{23}$Na reaction. During He-burning, \F{} is produced by $\alpha$-captures on $^{15}$N, and also depleted by $\alpha$-captures, producing \Ne{} \citep[][]{MeynetArnouldF192000}. \Ne{} is produced in the helium burning, mainly by double $\alpha$-captures on $^{14}$N, while it is not significantly destroyed. To expel a significant amount $^{19}$F into the interstellar medium, the He-core needs to be exposed in an early stage of helium-burning, before $^{19}$F is destroyed.\\
\indent For the majority of the models shown in Figure \ref{fig:Stables}, the initial abundance (red lines) of the stable isotopes are higher than the amount expelled from the stars, and therefore the net yield is negative. Only the heaviest, rotating models in our set ($\geq$ 60 \msun{}) produce a positive net yield. This is in agreement with the earlier results for \F{} by \cite{MeynetArnouldF192000} and \cite{PalaciosF192005}.\\
\indent For \Ne{} the stars have positive yields for masses 40-45 M$_{\odot}$ and above, depending on initial rotational velocity. This is because, unlike \F{}, \Ne{} is not completely destroyed during hydrogen burning and during helium burning, \Ne{} is not completely destroyed while \F{} does. Therefore, it is easier to obtain a positive yield. These yields are not only sensitive to the wind prescription, which determines how much material is ejected, but also to the internal mixing processes due to rotation, which leads to more or less destruction of the initial abundance.
\subsection{Comparison to other data sets}
The right panels of Figures \ref{fig:SLRs} and \ref{fig:Stables} show the results of our models along with several other studies from the literature. On top LC18 and E12, here we also consider the non-rotating models from \cite{Sukhbold2016}, computed with the KEPLER stellar evolution code, \cite{Pignatari2016}, computed with GENEC for the for mass range of interest, and from \cite{Ritter2018}, computed with MESA version 3709. We note that E12 and \cite{Pignatari2016} both use the GENEC code, however, the models of \cite{Pignatari2016} cover all burning stages, therefore all isotopes we are interested in are included, while for for E12 we do not have yields for \Cl{}, \Ca{}, and \Fe{}. Overall, the yield sets are in broad agreement within the uncertainties of the stellar evolution modelling. The models that stand out the most are the low-mass (10-30 \msun{}) LC18 models. These models show an increase in the SLR yields from non-rotating to rotating orders of magnitude higher than found by the other models. For \Al{}, the LC18 increase is comparable to the increase in the \Al{} yields we found in Paper\,I between single stars and binary models. This behaviour in the LC18 models is due to a very strong increase in the mass loss compared to their non-rotating models, leading to an increased yield. This difference in mass loss is due to the treatment of rotational mixing and the formation of a dust driven wind, as explained in \cite{LandC2013}.\\
\indent For the stable isotopes, also the highest mass models by \cite{Sukhbold2016} stand out. These models lose much more mass than our 60-80 \msun{} stars, especially their 80 \msun{} model loses nearly 74 \msun{}, compared to $\simeq$ 56 \msun{} for our non-rotating model. The 20 and 25 \msun{} E12 models have higher yields than the overall trend for these two masses, because these stars have a slight boost to their winds, independent of rotation.
\begin{table*}[t]
\caption{Isotopic yields in \msun{}. M$_{\rm ini}$ is the initial mass in M$_{\odot}$. V$_{\rm ini}$ is the initial rotational velocity in km/s. For the stable isotopes, the initial amount present in the star is also given. The yields tabulated here are not corrected for radioactive decay that might take place in the ISM during the evolution of the star. Yields for all 209 isotopes in the nuclear network are available on Zenodo: \url{https://doi.org/10.5281/zenodo.5497258}.}
\begin{center}\begin{tabular}{cc|cc|cc|ccc}
\hline
M$ _{\rm ini} $ &  V$ _{\rm ini} $& $^{19}$F$_{\rm ini}$ & $^{19}$F &$^{22}$Ne$_{\rm ini}$ & $^{22}$Ne&$^{26}$Al & $^{36}$Cl& $^{41}$Ca\\
(M$ _{\odot}$)& (km/s) &(M$ _{\odot}$) &(M$ _{\odot}$) &(M$ _{\odot}$) &(M$ _{\odot}$) &(M$ _{\odot}$) &(M$ _{\odot}$) &(M$ _{\odot}$)\\
\hline
10 & 0 &4.67e-06 &4.61e-07 &8.90e-4 &8.47e-05 &7.19e-11 &8.82e-23 &7.90e-23\\ 
 & 150 &4.67e-06 &4.71e-07 &8.90e-4 &8.68e-05 &2.78e-10 &1.11e-22 &1.29e-22\\ 
 & 300 &4.68e-06 &4.20e-07 &8.92e-4 &7.77e-05 &1.40e-10 &1.67e-22 &1.70e-22\\
\hline
15 & 0 & 6.42e-06 & 1.69e-06& 1.32e-3 & 3.12e-4 &8.85e-09 &3.00e-22 &1.41e-21\\
 & 150 & 6.42e-06 &1.74e-06 &1.32e-3 &3.24e-4 &1.47e-08 &5.12e-22 &2.10e-21\\
 & 300 & 6.45e-06 &1.21e-06 &1.33e-3 &2.26e-4 &2.34e-08 &5.32e-22 &3.06e-21\\
\hline 
20 & 0 &8.22e-06 &4.01e-06 &1.79e-3 &7.43e-4 &1.81e-07 &1.94e-21 &1.28e-20\\
 & 150 &8.24e-06 &4.04e-06 &1.79e-3 & 7.57e-4 &1.92e-07 &9.48e-22 &1.33e-20\\
 & 300 &8.28e-06 &2.04e-06 &1.79e-3 &3.86e-4 &5.87e-07 &1.75e-21 &2.87e-20\\
\hline
25 & 0 &9.67e-06 &5.87e-06 &2.23e-3 &1.09e-3 &1.17e-06 &2.80e-21 &6.27e-20\\
 & 150 &9.69e-06 &5.23e-06 &2.23e-3 &9.89e-4 &1.59e-06 &4.52e-21 &9.28e-20\\
 & 300 & 9.75e-06 &2.28e-06 &2.24e-3 &4.41e-4 &3.93e-06 &4.98e-16 &2.06e-15\\ 
\hline
30 & 0 &1.11e-05&6.38e-06 &2.69e-3 &1.19e-3 &3.41e-06 &6.76e-21 &2.22e-19\\
 & 150 & 1.12e-05 &5.84e-06 &2.70e-3 &1.12e-3 &4.68e-06 &3.30e-15 &1.42e-14\\
 & 300 &1.12e-05 &2.39e-06 &2.70e-3 &4.77e-4 &9.89e-06 &1.45e-13 &6.66e-13\\
\hline 
35 & 0 &1.33e-05 &6.77e-06 &3.20e-3 & 1.27e-3 & 8.44e-06&4.70e-16 & 1.96e-15\\
 & 150 &  1.33e-05 &5.84e-06 &3.20e-3 &1.13e-3 &1.06e-05 &6.85e-13 &3.23e-12\\
 & 300 &1.34e-05 &2.51e-06 &3.20e-3 &5.27e-4 &2.11e-05 &1.85e-10 &9.29e-10 \\
\hline 
40 & 0 & 1.42e-05 &7.06e-06 &3.64e-3 &1.36e-3 &1.88e-05 &7.99e-11 &4.015e-10\\
 & 150 & 1.43e-05 &6.14e-06 &3.64e-3 &2.13e-3 &2.05e-05 &3.56e-08 &6.07e-08\\
 & 300 & 1.44e-05 &2.82e-06 &3.65e-3 &5.09e-3 &3.43e-05 &1.82e-07 &2.99e-07\\
\hline
45 & 0 & 1.61e-05 &7.40e-06 &4.13e-3 &5.18e-3 &2.94e-05 &1.45e-07 &2.44e-07\\
 & 150 &1.61e-05 &6.31e-06 &4.13e-3 &1.01e-2 &3.15e-05 &3.59e-07 &5.86e-07\\
 & 300 & 1.63e-05 &3.23e-06 &4.13e-3 &1.54e-2 &5.32e-05 &5.99e-07 &9.78e-07\\
\hline
50 & 0 &1.75e-05  &7.82e-06 &4.60e-3  &1.21e-2  &4.00e-05  &4.18e-07  &6.89e-07\\
 & 150 &1.76e-05 & 7.00e-06 &4.60e-3 &2.30e-2 &4.30e-05 &8.78e-07 &1.43e-06\\
 & 300 & 1.77e-05 &3.42e-06 &4.61e-3 &2.14e-2 &7.04e-05 &8.26e-07 &1.34e-06\\
\hline 
60 & 0 &1.91e-05 &9.00e-06 &5.50e-3  &3.84e-2 &6.65e-05  &1.48e-06 &2.42e-06\\
 & 150 & 1.91e-05 &7.28e-06 &5.50e-3 &4.23e-2 &7.16e-05 &1.63e-06 &2.67e-06\\
 & 300 & 1.93e-05 &1.91e-05 &5.51e-3 &0.12 &1.44e-4 &3.88e-06 &8.37e-06\\
\hline
70 & 0 & 2.09e-05 &9.21e-06 &6.43e-3 &5.58e-2 &9.70e-05 &2.10e-06 &3.46e-06\\
 & 150 &2.10e-05 &1.78e-05 &6.43e-3&1.38e-1 &1.25e-4 &5.00e-06 &9.55e-06\\
 & 300 &2.11e-05 &3.11e-05 &6.43e-3 &0.17 &2.36e-4 &4.48e-06 &1.14e-05\\
\hline
80 & 0 & 2.43e-05 &1.49e-05 &7.41e-3 &0.14 &1.51e-4 &5.40e-06 &9.55e-06\\
 & 150 &2.44e-05 &5.16e-05 &7.41e-3 &0.18 &2.69e-3 &4.52e-06 &1.20e-05\\
 & 300 &2.46e-05 &3.51e-05 &7.41e-3 &0.11 &4.29e-4 &2.59e-06 &6.99e-06\\
\hline
\end{tabular}\end{center}
\label{Yields}
\end{table*}
\section{Early Solar System}\label{sec:ESS}
The radioactive isotopes we have studied in the previous section were inferred to be present in the early Solar System (ESS) from observed excesses of their daughter nuclei in meteoritic inclusions. In this section, we consider a simple dilution model for \Al{}, \Cl{}, and \Ca{}, to investigate if their abundances in the ESS can be explained self-consistently with the models presented here. While other stellar objects in the galaxy can produce these isotopes, such as novae and asymptotic giant branch stars, much research has focused on massive stars as ESS polluters, because these stars live short enough to be able to eject material within star forming regions which have typical lifetimes of at most a few tens of Myr \citep[][]{Murray2011}. Furthermore, massive star winds are preferred by several authors (\citealt{Gaidos2009,Gounelle2012,Young2014,Dwarkadas2017}) as a most favoured site of origin for \Al{}, because there are several difficulties for core-collapse supernovae to produce the abundances of SLRs in the proportion required to match the ESS values \citep[see discussion in][]{Lugaro2018}.
For example, they produce too high abundances of \Fe{} and $^{53}$Mn. Both these isotopes have half lives 4 to 50 times higher than the SLRs that we consider in the calculation here, therefore, their abundances in the ESS can be explained by decaying their abundances in the interstellar medium as derived by galactic chemical evolution (see, e.g., \citealt{Wasserburg2006, TangDauphas2012, Trappitsch18, Cote2019GCESLRs,Cote2019MonteCarlo}).
\subsection{The dilution method}
\label{delaytime}
The comparison to the ESS values is performed in four steps:\\
\indent Step 1 is to determine a ``dilution factor", $f$ for each stellar model. This is defined as $f=\frac{M^{\rm ESS}_{\rm SLR}}{M^{*}_{\rm SLR}}$, where $M^{\rm ESS}_{\rm SLR}$ is the mass of a given SLR in the ESS, and $M^{*}_{\rm SLR}$ is the mass of the same SLR ejected by the stellar wind, i.e., the total yield. We use \Al{} to determine f and then apply the same value to \Cl{} and \Ca{}\footnote{If the injection of the SLRs occurred in the form of dust grains, chemistry and dust formation could lead to different $f$ values for the different isotopes because Al, Ca, and Cl are different elements. While Al and Ca should behave chemically in a very similar way, the situation for Cl may be different. We do not consider these uncertainties here.}, this is because the \Al{}/$^{27}$Al ratio in the ESS is very well established to $5.23 \times 10^{-5}$, as reported by \citet{Jacobsen2008} and recently confirmed by \cite{Luu2019}. We derive an initial amount of \Al{} of 
3.1$\times$10$^{-9}$\msun{} assuming the solar abundance of $^{27}$Al \citep{Lodders2003} and a total mass of 1 \msun{} to be polluted \citep[see details in][]{Lugaro2018}.\\
\indent Step 2 is to determine the ``delay time" ($\Delta$t), which is the time interval between wind ejection and the formation of the first solids (the calcium-aluminium rich inclusions (CAIs)) in the ESS, for which time the SLR values are given. After we have applied our determined $f$ to the yields of \Ca{}, to obtain the diluted amount of this isotope in Step 1, we decay the \Ca{} until we reach the observed ESS \Ca{}/$^{40}$Ca ratio of 4.6$\times$10$^{-9}$. The time needed, is the delay time.\\
\indent Step 3 is to calculate a new abundance of \Al{} by reverse decaying the initial ESS amount of \Al{} using the delay time from Step 2. With this we recalculate $f$ and use it to repeat Step 2. We continue this iteration until we converge to a $\Delta$t value within a 10\% difference from the previous value.\\
\indent Step 4 apply the final $f$ to calculate the diluted \Cl{} abundance. We then repeat Step 2 for \Cl{} also to determine a delay time for \Cl{} ($\Delta$t$_{Cl}$) using the ESS \Cl{}/$^{35}$Cl ratio of 2.44$\times$10$^{-5}$ measured in the \textit{Curious Marie} CAI \citep[][]{Tang2017}.\\\\
\indent A few remarks need to be made about this method. First, even though \Al{} is not produced in the same evolutionary phase as \Cl{} and \Ca{}, the bulk of these isotopes are expelled into the interstellar medium at the same time. Therefore, we do not need to take into account, for example, that some \Al{} might have decayed before \Cl{} and \Ca{} were ejected the star. Second, we use the \Ca{}/$^{40}$Ca-ratio to obtain $\Delta$t even though this ratio is not well constrained in the ESS \citep{Liu2017}, because \Ca{} is the shortest lived of the three isotopes considered here. Therefore, it is the most sensitive chronometer to short timescales \citep[see also][]{Wasserburg2006}. Moreover, the abundances of both \Cl{} and \Ca{} can have a contribution from irradiation by solar cosmic rays in the disk. Higher values of \Cl{} than the ESS value used here have been measured in other meteoritic inclusions and can be produced by irradiation within the ESS. The value measured in \textit{Curious Marie} probably represents the primordial value derived from a stellar source because \Cl{} coexists in this CAI with the canonical value of \Al{} \citep{Tang2017}. We note also that the measurements of the ESS values of \Ca{} and \Cl{} might be affected by systematic uncertainties, because the abundance of \Ca{} is very low and the abundance of \Cl{} is based on the measurement of its roughly 2\% decay channel into $^{38}$S. For \Ca{} the latest data on a handful of CAIs \citep{Liu2017} demonstrate the presence of this very short-lived isotope in the ESS, however, the data precision is not high enough to be able to resolve possible heterogeneities.
\subsection{Results and comparison to other studies}
\begin{figure*}
    \begin{center}
    \includegraphics[trim = 5mm 1mm 5mm 13mm, width=0.49\textwidth]{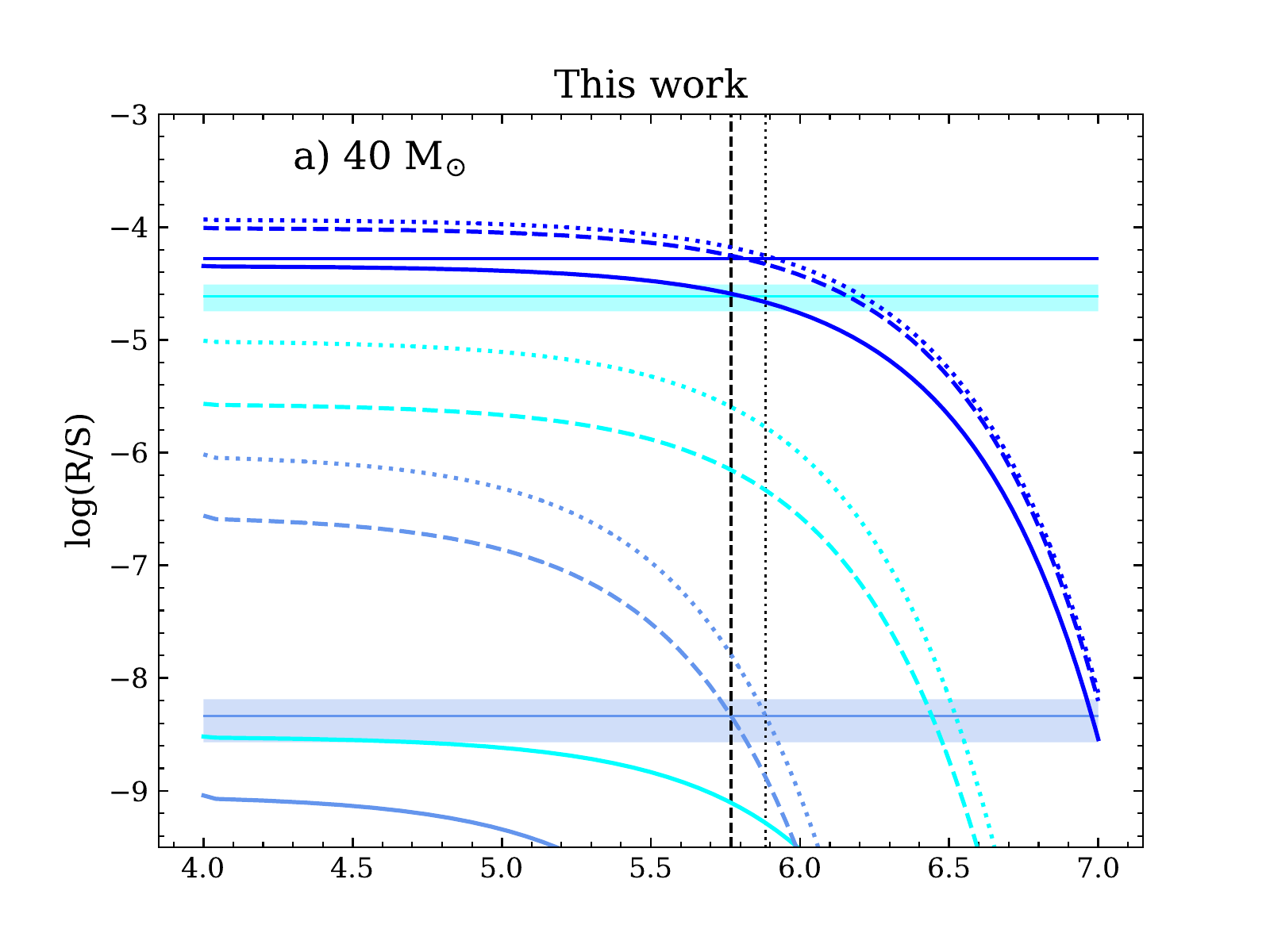}
    \includegraphics[trim = 5mm 1mm 5mm 13mm, width=0.49\textwidth]{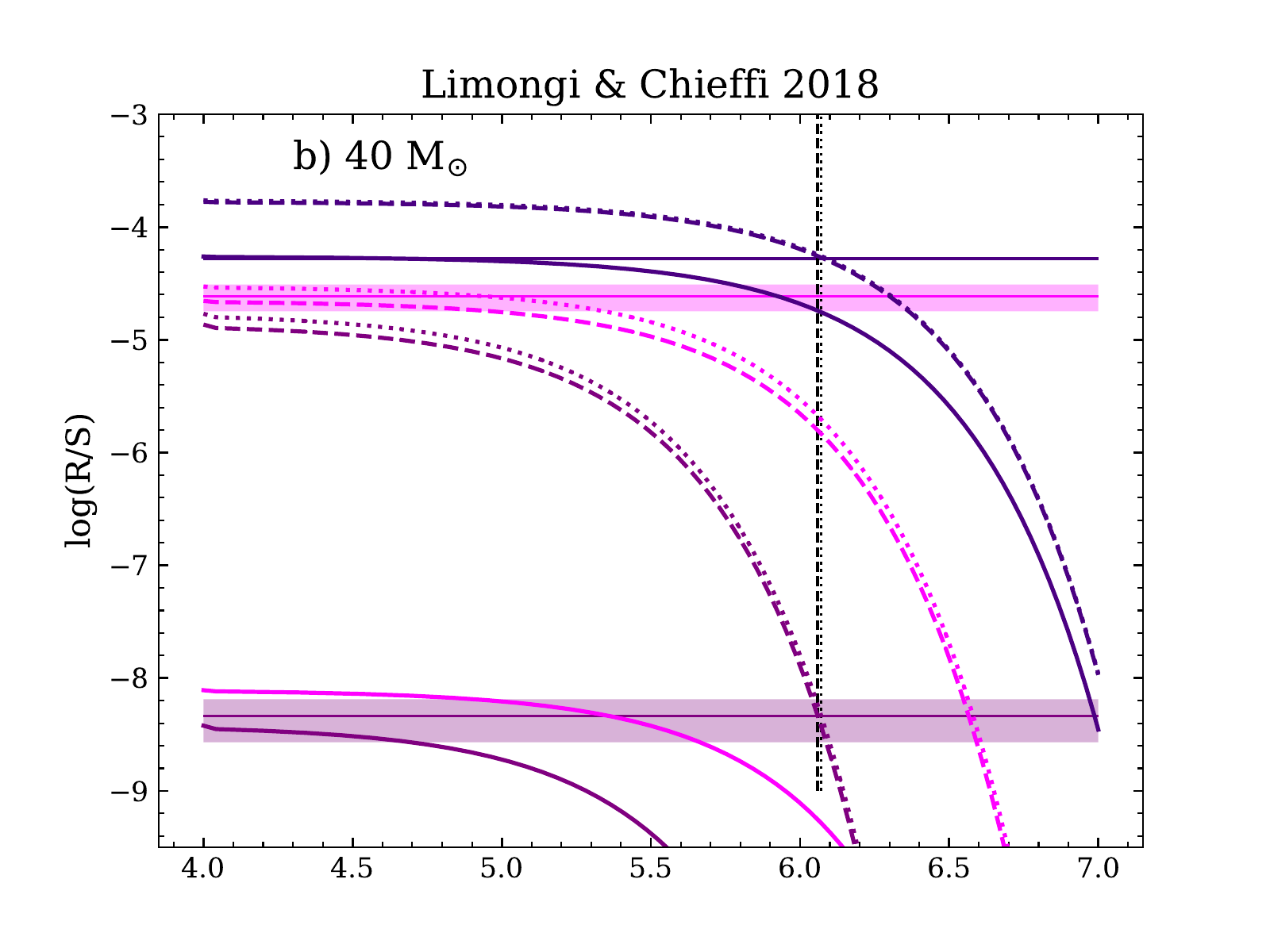}\\
    \includegraphics[trim = 5mm 1mm 5mm 13mm, width=0.49\textwidth]{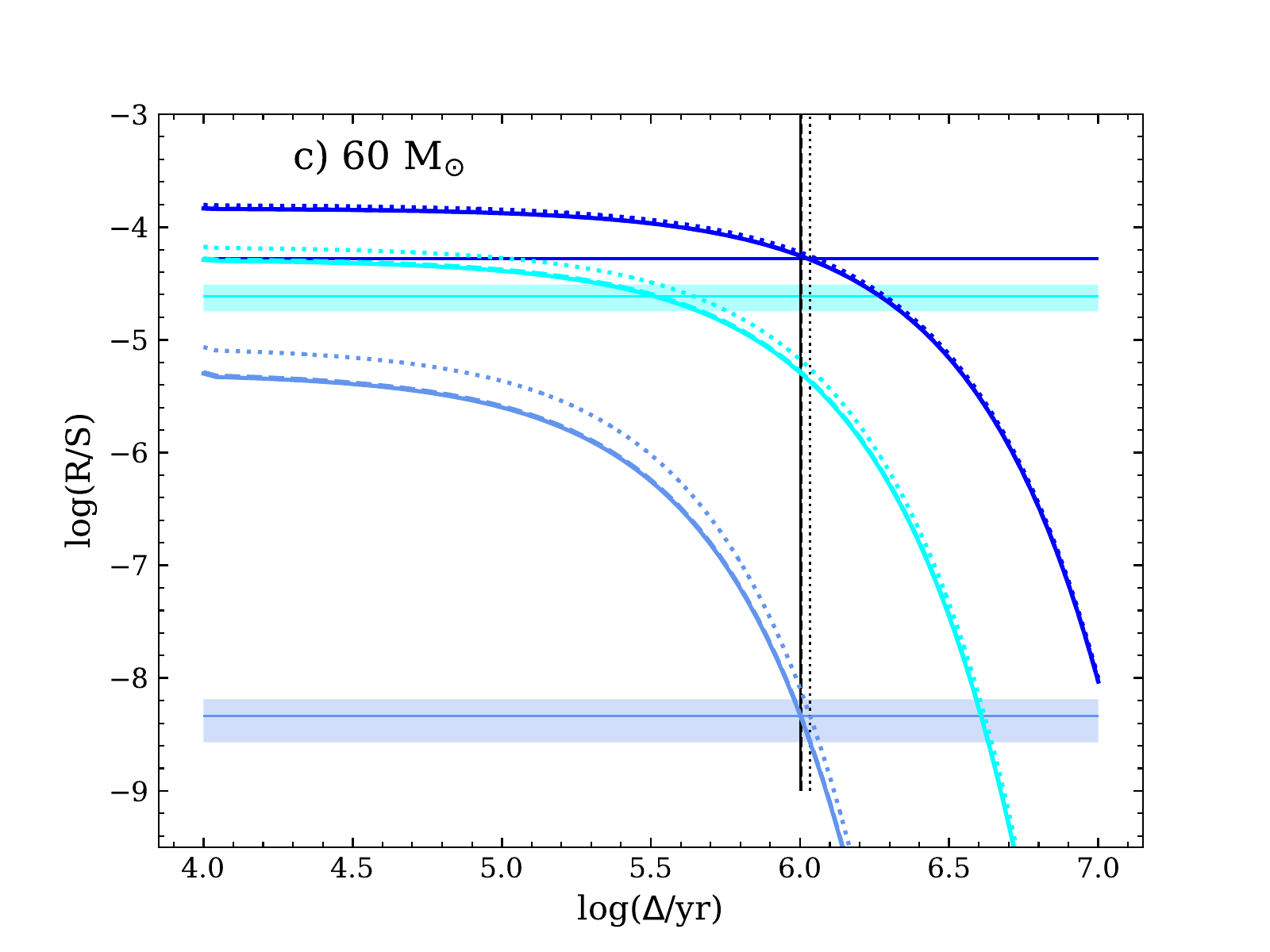}
    \includegraphics[trim = 5mm 1mm 5mm 13mm, width=0.49\textwidth]{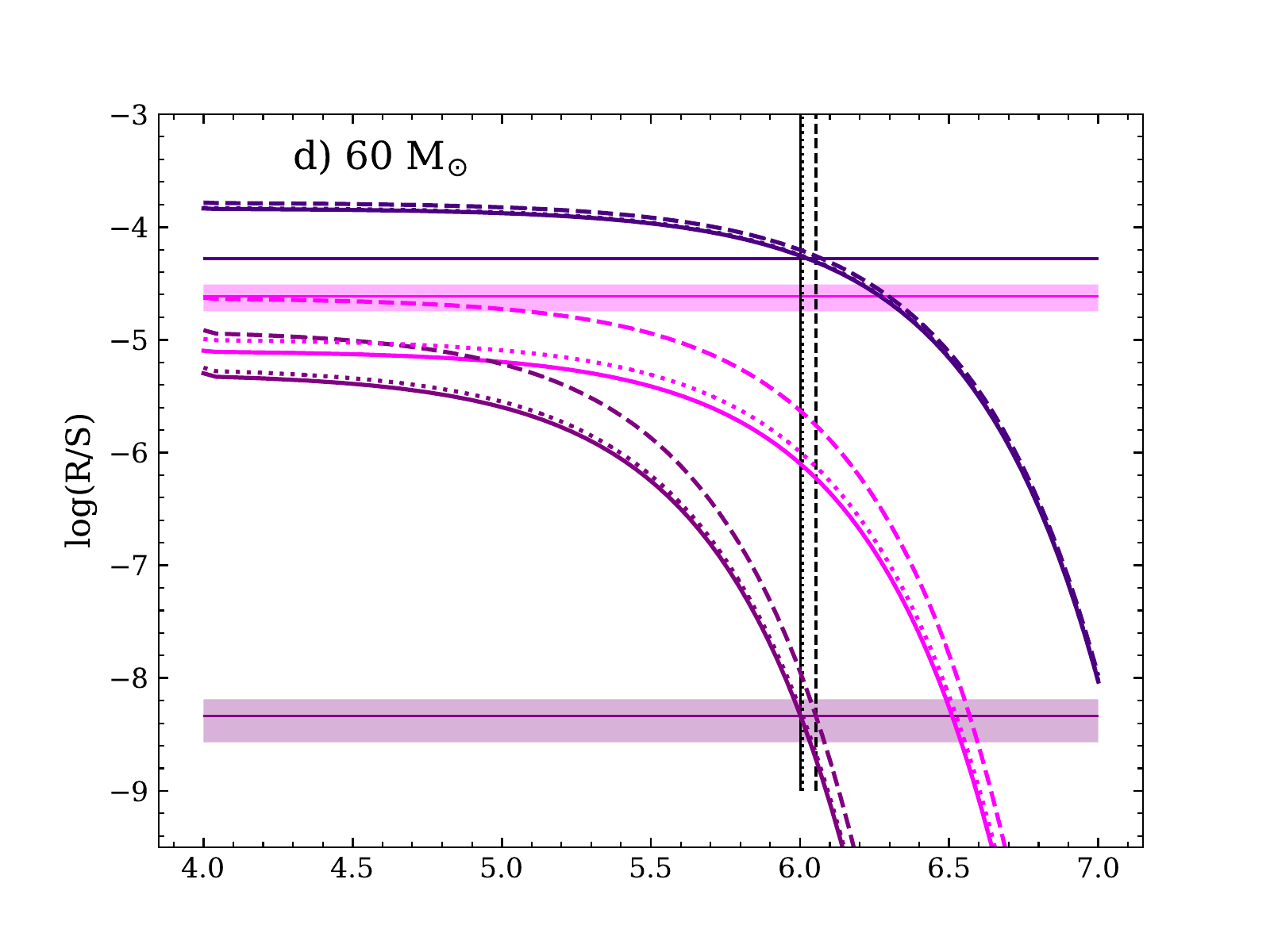}\\
    \includegraphics[trim = 5mm 1mm 5mm 13mm, width=0.49\textwidth]{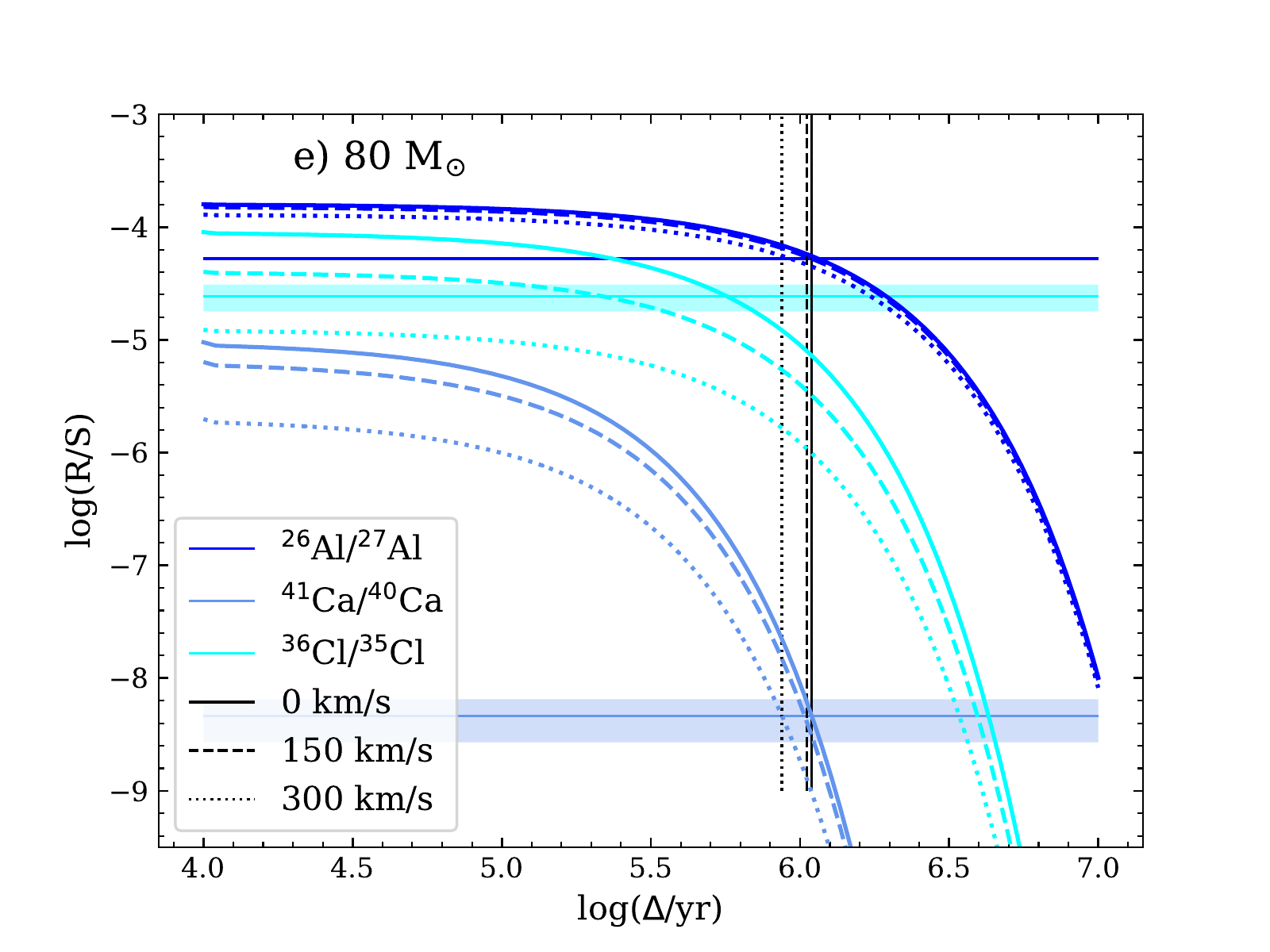}
    \includegraphics[trim = 5mm 1mm 5mm 13mm, width=0.49\textwidth]{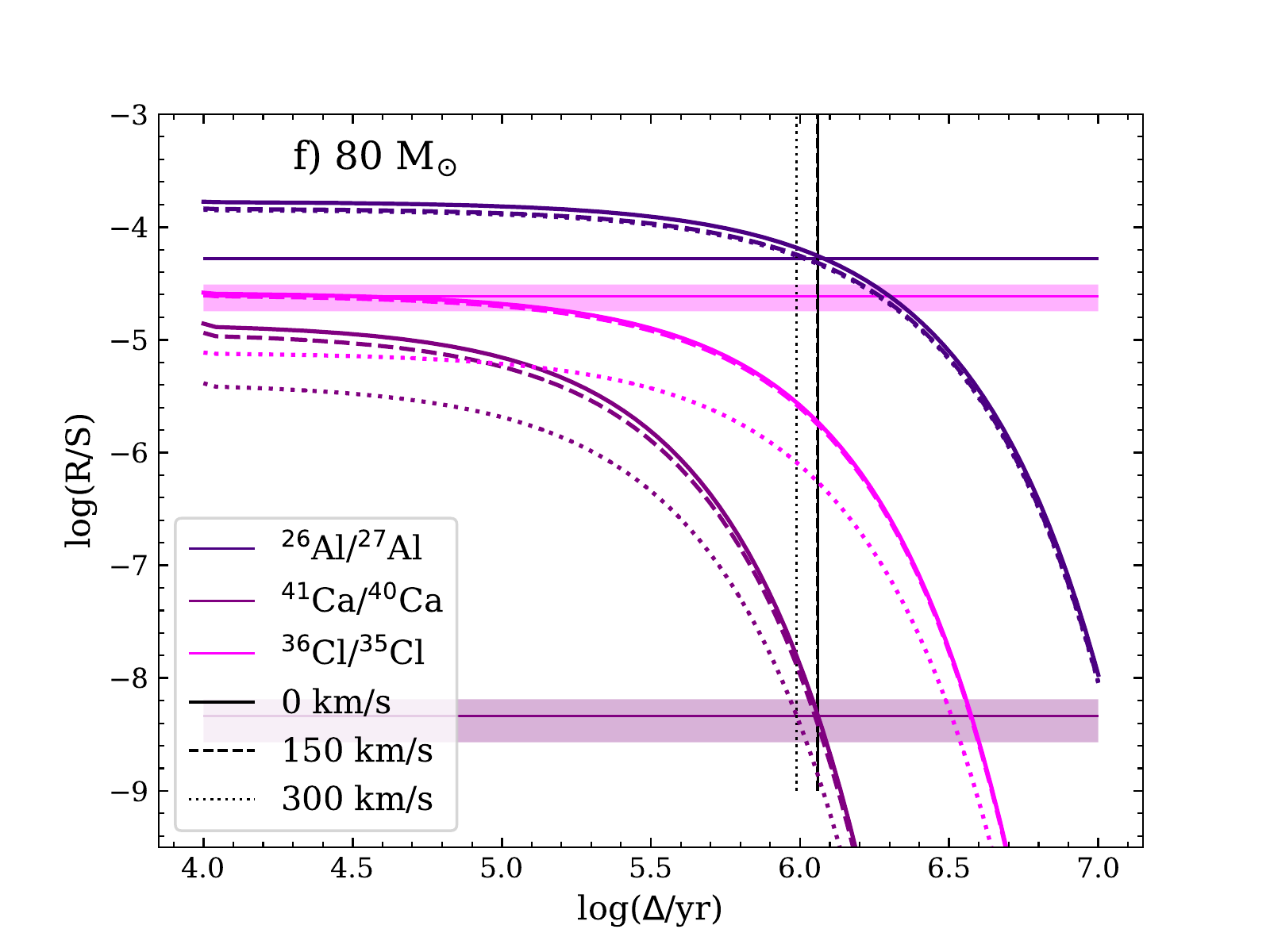}
    \end{center} 
    \caption{Abundance ratios (R/S) for the three SLRs (R) over their stable reference isotope (S) for a certain dilution factor $f$. The solid lines represent the non-rotating models, the dashed lines the models with an initial rotational velocity of 150 km/s, and the dotted lines the models with an initial rotational velocity of 300 km/s. The horizontal bands represent the ESS ratios, with their respective errors. The vertical lines represent the delay time for the Ca-ratio. The left panels give the results for our models (40, 60, and 80 \msun{}), the right panels give the results for the models by LC18 for the same masses. The range of values of f for each panel (in units of 10$^{-4}$) are: a) 1.32-2.64, b) 2.44-3.03, c) 0.60-1.13, d 0.54-1.37, e) 0.16-0.5, and f) 0.08-0.24.} 
    \label{fig:DelayTime}
\end{figure*}

We apply the method described in Sec. \ref{delaytime} to our models and they are considered to be a solution for the ESS when the delay times for \Ca{} and \Cl{} are comparable with a factor of up to 5. For our non-rotating models, a solution can be found for \Al{} and \Ca{} in the mass range 45 to 80 \msun{}
and $\Delta$t between 0.7 and 1 Myr. In order to also match the \Cl{} abundance, the mass-range needs to be restricted to stars with an initial mass of 60 \msun{} and higher. There still is a small inconsistency, because $\Delta$t needs to be lower for \Cl{} than for \Ca{}, as the delay time for \Cl{} is between 0.2 and 0.5 Myr. The results for our rotating models are similar, except that the mass range for which a solution is possible for \Al{} and \Ca{} may extend down to 40 \msun.
If we consider the three isotopes, it is possible to find a solution for initial rotational velocity of 150 and 300 km/s at masses 50-80 \msun{} and 60 and 70 \msun{}, respectively.\\
\indent In Figure \ref{fig:DelayTime} we look more closely at a selected set of models. It shows the abundance ratios (R/S) for the three SLRs (R) over their stable reference isotope (S) versus delay time for the dilution factor $f$, for our stellar models with a mass of 40, 60, and 80 \msun{}. The horizontal bars represent the ESS-ratios and their uncertainties. We compare our results with the results calculated from the LC18 models for the same initial masses.\\
\indent For the 40 \msun{} models (Figure \ref{fig:DelayTime}a), the \Cl{} and \Ca{} yields from the non-rotating model (solid lines) are too low to match their ESS ratios.
Only our rotating models (dashed and dotted lines in panel a) can match both the Al- and Ca-ratios (at 10$^{5.9}$ yrs), while the the Cl-ratio is between one and two orders of magnitude too low.
This excludes this star as a potential solution for the ESS. For the LC18 models (Figure \ref{fig:DelayTime}b), the non-rotating model does not match Cl- or the Ca-ratio either.
As for our models, the rotating models match both the Al- and Ca-ratios. The difference is that in the LC18 models the Cl-ratio is matched very early on in the calculation, at 10$^{4.9}$\,yrs, while the other ratios are matched after 1 Myr, while in our models the Cl-ratio cannot be matched at all. However, due to this large difference in $\Delta$t, these models can still not be considered a potential solution.\\
\indent For the 60 \msun{} models (Figure \ref{fig:DelayTime}c), all of our models can match the three isotopic ratios, the Cl-ratio at 10$^{5.5}$\,yrs and the other two ratios at 10$^{6}$\,yrs. For the LC18 models (Figure \ref{fig:DelayTime}d), only their model with an initial rotational velocity of 150 km/s matches the Cl-ratio, but too early (10$^{4-4.6}$\,yrs) to be considered a solution together with the Al- and Ca-ratios.\\
\indent Finally, for the 80 \msun{} (Figure \ref{fig:DelayTime}e), we can match the ESS ratios of the SLRs with our non-rotating model and the model rotating at an initial velocity of 150 km/s. There is still a difference in the delay times for \Ca{} and \Cl{} of a factor of 2. For the 80 \msun{} LC18 models (Figure \ref{fig:DelayTime}f), only the non-rotating model matches all three ratios, and the Cl-ratio is matched at 0.03 Myr, while the other two are matched at 1.14 Myr. This difference is again too large to consider this as a solution. The rotating models cannot match the Cl-ratio.\\
\indent The difference between the sets can be explained by considering Figures \ref{fig:SLRs}d and \ref{fig:SLRs}f.
While the \Ca{} yields are very comparable between the two sets of both non-rotating and rotating models for 60 and 80 \msun{}, the non-rotating and rotating \Cl{} yields are lower for the LC18 models due to the differences in mass loss (see Section \ref{slrs}) compared to our yields. Therefore, there is less \Cl{} compared to \Ca{} in the LC18 models than in our models, making it harder to match both ratios within the uncertainties. For the 40 \msun{} models, the rotating yields given by LC18 are higher than our rotating yields for \Cl{}, and \Ca{} which explains why their rotating models can match for the Cl-ratio, while our models can not.\\
\indent Our results that the three radionuclides \Al{}, \Cl{}, and \Ca{} can be ejected by the winds of a variety of WR stars at relative levels compatible with the meteoritic observations are also in qualitative agreement with the results of \cite{Arnould1997,Arnould2006}. Because these authors used $^{107}$Pd to calculate $f$, our results are not directly comparable. However, if we consider the delay times shown in Figures 5-7 of \cite{Arnould2006}, we find similar values for the delay time for the Ca-ratio, between 10$^{5.5}$-10$^{6}$\,yrs. Their delay time for the Cl-ratio for their 40 \msun{} model is much closer to their delay time for the Ca-ratio compared to our models and those by LC18. However, the delay times for the Cl-ratio are worse for their 60 and 85 \msun{} models than for our models or those by LC18. Our result also confirm the analysis of the production of \Al{} in WR stars of mass between 32 and 120 \msun{} by \cite{Gounelle2012}. Our $f$ factor may be compared to their $\eta_{\rm wind}$/1000 in their Equation~(2), for which they find values ranging down to 2$\times$10$^{-5}$. For our models it is possible to find a solution for \Al{} and \Ca{} for models with initial masses 60-80 \msun{} with dilution factors in the range 0.00011 to 7.2 $\times$ 10$^{-5}$ and $\Delta$t around 1 Myr.\\
\indent Overall, we conform that WR stars are a robust candidate site for the production of \Al{}, \Cl{}, and \Ca{} in the ESS.
\subsection{The impact of neutron-capture rates}
\begin{figure}
   \centering
    \includegraphics[width=0.49\textwidth]{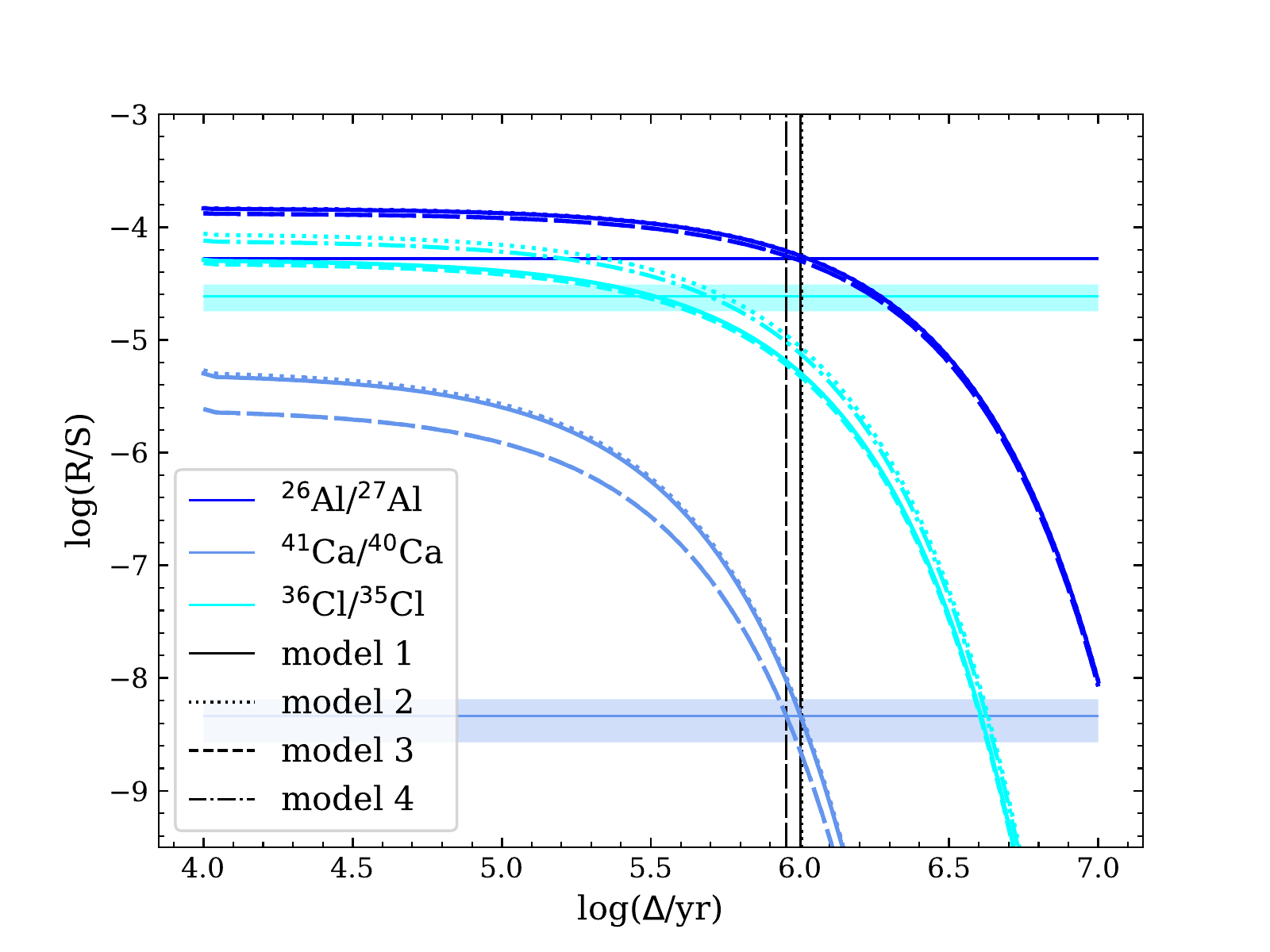}
    \caption{Results for the reaction rate tests. The colours and bands are the same as in Figure \ref{fig:DelayTime}.}
    \label{fig:NeutronCaptures}
\end{figure}

From Figure \ref{fig:DelayTime}, it becomes clear that to match the Al-, Ca-, and Cl-ratios using a self-consistent delay time within the scenario and the ESS ratios discussed in this work, more \Cl{} and/or less \Ca{} is required. To look more closely at this, we performed a sensitivity study for the destruction of these two isotopes. The dominant rate in the \Cl{} and \Ca{} destruction via neutron-captures are the \Cl{}(n,p)$^{36}$S and \Ca{}(n,$\alpha$)$^{38}$Ar rates, respectively. For the rate \Cl{}(n,p)$^{36}$S we use the reaclib label ``ths8'' which corresponds to a theoretical determination by \citet{RauscherNonSmoker}. For \Cl{}(n,p)$^{36}$S reported experimental rates are both similar to the value used here \citep{deSmet2007} and 30\% lower, at the temperature of interest here around 200-300 MK. For the  \Ca(n,$\alpha$)$^{38}$Ar rate the JINA reaclib reference is \citet{SEVIOR1986128}, which is roughly a factor of two higher than the theoretical rate by \citet{RauscherNonSmoker}. Given that there are significant differences between the current estimates, we computed three additional 60 \msun{} non-rotating models for which we multiplied the neutron-capture reaction rates of interest by different constants as indicated in Table \ref{tab:tableClCamodeltest}.\\
\indent The table also gives the yields for \Cl{} and \Ca{}. The \Cl{} yield increases when the reaction rate is decreased, as was expected (models 2 and 4). For model 2, the \Ca{} yield increases slightly.
For models 3 and 4 the \Ca{} yield decreases by about a factor 2, as is expected as well.
The changes in the yields are not fully linear due to other (neutron-capture) reactions also playing a minor role in the destruction of the these two isotopes. If we apply the delay time calculation to these new models, we find that for model 1, the difference in the delay time between the Ca-ratio and the Cl-ratio is a factor of $\sim$3. For model 2, the delay time difference is similar to that of model 1. This is visible in Figure \ref{fig:NeutronCaptures}, where the solid lines (model 1) and the dashed lines (model 2) are more or less overlapping. Models 3 (dotted lines) and 4 (dashed-dotted lines), provide a better solution as the difference in the delay time is reduced to a factor 1.8. There is little difference between these two models. From this, we can conclude that to obtain a better match for all three SLR-ratios, a decrease in the amount of \Ca{} has more impact than the increase in the amount of \Cl{}. 
\begin{table}
        \caption{Factors used to multiply the indicated reaction rates from their standard values in the 60 M$_{\odot}$ non-rotating models considered in this section and the yields in \msun{} resulting from these changes.}
    \label{tab:tableClCamodeltest}
    \begin{tabular}{ccc|cc}
      & $^{36}$Cl(n,p)$^{36}$S & $^{41}$Ca(n,$\alpha$)$^{38}$Ar&\Cl{} (\msun{}) &\Ca{} (\msun{})\\
      \hline          
      Model 1 & 1   & 1 & 1.48e-06 &2.42e-06\\
      Model 2 & 0.5 & 1  & 2.47e-06 &2.55e-06\\
      Model 3 & 1   & 2 & 1.51e-06 &1.29e-06\\
      Model 4 & 0.5 & 2 & 2.41e-06 &1.29e-06\\
      \hline
    \end{tabular}
\end{table}

\subsection{Oxygen isotopic ratios}
\indent For sake of completeness, we also checked if the models that reproduce \Al{} and \Ca{} affect the O isotopic ratios. This is because a successful pollution model should avoid predicting a correlation between the presence of \Al{} and modification of the O isotopes \citep{GounelleMeibom2007} to match the observational evidence that some CAIs (the Fractionated and Unknown Nuclear anomalies, FUN, class) and some corundum grains are poor in \Al{}, but they have virtually the same O isotopic composition as those that are rich in \Al{} \citep[see, e.g.,][]{Makide2011}. 
The composition of the O isotopes in the winds is dominated by H and He burning and its main features are production of $^{16}$O and depletion of $^{17}$O and $^{18}$O, relative to their initial amounts in the star, with $^{17}$O/$^{16}$O and $^{18}$O/$^{16}$O ratios at most roughly 8 and 20 times higher than solar. Therefore, when we add the O isotopic wind yields of the selected models diluted by $f$ to the O solar system abundances, we obtained a decrease of the order of 0.1 to 1.5\% in both the $^{17}$O/$^{16}$O and $^{18}$O/$^{16}$O ratios, increasing the absolute value with increasing the stellar mass from 40-50 \msun{} to 80 \msun{}. While variations at the lower values of this range would not be detectable, those at the upper values would be. Therefore, if we assume the specific scenario where the \Al{}-poor and \Al{}-rich grains formed before and after injection, respectively, then the lower masses would be favoured to avoid changes in the O ratios. However, the higher masses reproduce the \Cl{} abundance. This discrepancy needs to be further investigated.

\section{Conclusion}\label{sec:Conclusion}
We have investigated the production of selected stable and radioactive isotopes in the winds of massive rotating stars. The selected SLRs of interest for the ESS are: \Al{}, \Cl{}, and \Ca{}, while the selected stable isotopes are: \F{} and \Ne{}. These results have been compared to various studies in the literature. We also determined which of these models could self-consistently explain the ESS abundances of \Al{}, \Cl{}, and \Ca{}. We have found that:
\begin{itemize}
    \item For the SLRs, it is mostly the WRs in the mass-range 40-80 \msun{} that give significant yields. However, the \Fe{} yields are insignificant compare to the supernova yields.
    \item Only our most massive rotating stars produce a net positive \F{} yield ($\geq$60 \msun{} for an initial rotation rate of 300\,km/s). For \Ne{}, more stars give a net positive yield, from $\geq$40-45 \msun{}, depending on the initial rotational velocity. 
    \item The main effect of rotation is that it lowers the initial mass for which the stars become WRs. For \Al{} the effect of rotation on the yields is minimal, and only noticeable around the WR limit. For \Cl{} and \Ca{} a higher rotation rate leads to an increase in the yields at lower masses, shifting from $\sim$30 \msun{} to $\sim$20 \msun{}. From $\sim$45 \msun{} the yields become again comparable for all models. For the stable isotopes, the rotational mixing leads to lower yields below 50 \msun{} and 35 \msun{} for \F{} and \Ne{}, respectively.
    \item Overall, the yields from our models compare well to those from the literature. There are some differences caused by a different prescription of the mass loss and/or a different approach to rotational mixing. This clearly shows that the treatment of stellar winds and the increase of mass-loss due to rotation, as well as the treatment of rotation and rotational mixing, have still a large impact on the yields.
    \item In Section \ref{sec:ESS}, we have investigated which of the stellar models described in this paper could explain the ESS abundances. Depending on the initial rotational velocity, stars with an initial mass of 40-45 \msun{} and higher could explain the \Al{} and \Ca{} abundances. However, only the most massive models ($\geq$60 \msun{}) can also explain the \Cl{} abundances. We remind that also the following CCSNe of massive star models will expel a significant amount of these isotopes, however, they produce an overabundance of \Fe{} relatively to its ESS value.
    \item From our tests of the neutron-capture rates of \Cl{} and \Ca{} we conclude that to obtain a better match for all three SLR-ratios, a decrease in the amount of \Ca{}, derived from increasing its (n,$\alpha$) rate, has more impact than the increase in the amount of \Cl{}, derived from decreasing its (n,p) rate.
    \item When comparing our models with the oxygen-ratios in the Solar System, which are known to high precision, however, we find that the high mass models decrease the oxygen isotopic ratios, $^{17}$O/$^{16}$O and $^{18}$O/$^{16}$O, too much, while the lower mass models stay within the error-margins of the measurements.
\end{itemize}
\indent For all isotopes discussed here, except for \Al{} which we presented in Paper\,I, the influence of binary interactions still needs to be investigated. We will examine this in our upcoming paper, where we will also calculate whether these binary yields could match the ESS abundances. Future work includes the formation of dust from the WR binary stars \citep[see, e.g.,][]{Lau2020Dust}, since dust may be needed to incorporate SLRs into the ESS \citep{Dwarkadas2017}. Furthermore, a more detailed analysis of several uncertainties should be performed, which includes different prescriptions for the winds and for the rotational boost on wind loss, as well as investigations of the effect of reaction rate uncertainties specifically on the destruction of \F{}, \Ca{}, and \Cl{}, and the neutron source $^{22}$Ne($\alpha$,n)$^{25}$Mg reaction \citep{2021PhRvCAdsley}. Finally, to present a complete view of the isotopes discussed here, the explosive nucleosynthetic yields will need to be calculated using our models as the progenitors of the explosion.

\section*{Acknowledgements} \label{sec:thanks}
HEB thanks the MESA team for making their code publicly available, Frank Timmes for his support with the pre-supernova models, and Bill Paxton for other clarifications of the code. We thank Zsolt Keszthelyi for his help with implementing the rotational boost into MESA. We thank Alessandro Chieffi and Marco Limongi for clarifications of their results in their 2018 paper. We also thank Alan Boss for discussion on the ESS, and Andr\'es Yague for his assistance with Python.  We also thank the referee for all their comments and help with improving this paper. This work is supported by the ERC via CoG-2016 RADIOSTAR [11] (Grant Agreement 724560). We also acknowledge support from “ChETEC” COST Action(CA16117), supported by COST (European Cooperation in Science and Technology). MP acknowledges significant support to NuGrid from STFC (through the University of Hull's Consolidated Grant ST/R000840/1), from 
the National Science Foundation (NSF, USA) under grant No. PHY-1430152 (JINA Center for the Evolution of the Elements) and from the ``Lendulet-2014" Program of the Hungarian Academy of Sciences (Hungary). MP thanks the support from the US IReNA Accelnet network.
\software{MESA (\citealt{MESA1, MESA2, MESA3, MESA4})}

\bibliographystyle{apj}
\bibliography{references}

\begin{thebibliography}{}
\expandafter\ifx\csname natexlab\endcsname\relax\def\natexlab#1{#1}\fi

\bibitem[{{Abia} {et~al.}(2019){Abia}, {Cristallo}, {Cunha}, {de Laverny}, \&
  {Smith}}]{Abia2019}
{Abia}, C., {Cristallo}, S., {Cunha}, K., {de Laverny}, P., \& {Smith}, V.~V.
  2019, \aap, 625, A40

\bibitem[{{Adams}(2010)}]{adams10}
{Adams}, F.~C. 2010, \araa, 48, 47

\bibitem[{{Adsley} {et~al.}(2021){Adsley}, {Battino}, {Best}, {Caciolli},
  {Guglielmetti}, {Imbriani}, {Jayatissa}, {La Cognata}, {Lamia}, {Masha},
  {Massimi}, {Palmerini}, {Tattersall}, \& {Hirschi}}]{2021PhRvCAdsley}
{Adsley}, P., {Battino}, U., {Best}, A., {et~al.} 2021, \prc, 103, 015805

\bibitem[{{Aerts} {et~al.}(2019){Aerts}, {Mathis}, \& {Rogers}}]{Aerts2019}
{Aerts}, C., {Mathis}, S., \& {Rogers}, T.~M. 2019, \araa, 57, 35

\bibitem[{{Arnould} {et~al.}(2006){Arnould}, {Goriely}, \&
  {Meynet}}]{Arnould2006}
{Arnould}, M., {Goriely}, S., \& {Meynet}, G. 2006, \aap, 453, 653

\bibitem[{{Arnould} {et~al.}(1997){Arnould}, {Paulus}, \&
  {Meynet}}]{Arnould1997}
{Arnould}, M., {Paulus}, G., \& {Meynet}, G. 1997, \aap, 321, 452

\bibitem[{{Asplund} {et~al.}(2009){Asplund}, {Grevesse}, {Sauval}, \&
  {Scott}}]{Asplund2009}
{Asplund}, M., {Grevesse}, N., {Sauval}, A.~J., \& {Scott}, P. 2009, \araa, 47,
  481

\bibitem[{{Austin} {et~al.}(2017){Austin}, {West}, \& {Heger}}]{AustinSNe}
{Austin}, S.~M., {West}, C., \& {Heger}, A. 2017, \apjl, 839, L9

\bibitem[{{Basunia} \& {Hurst}(2016)}]{Alhalflife}
{Basunia}, M., \& {Hurst}, A. 2016, Nuclear Data Sheets, 134, 75

\bibitem[{{Belczynski} {et~al.}(2020){Belczynski}, {Klencki}, {Fields},
  {Olejak}, {Berti}, {Meynet}, {Fryer}, {Holz}, {O'Shaughnessy}, {Brown},
  {Bulik}, {Leung}, {Nomoto}, {Madau}, {Hirschi}, {Kaiser}, {Jones}, {Mondal},
  {Chruslinska}, {Drozda}, {Gerosa}, {Doctor}, {Giersz}, {Ekstrom}, {Georgy},
  {Askar}, {Baibhav}, {Wysocki}, {Natan}, {Farr}, {Wiktorowicz}, {Coleman
  Miller}, {Farr}, \& {Lasota}}]{BelczynskiRotation2020}
{Belczynski}, K., {Klencki}, J., {Fields}, C.~E., {et~al.} 2020, \aap, 636,
  A104

\bibitem[{{Brinkman} {et~al.}(2019){Brinkman}, {Doherty}, {Pols}, {Li},
  {C{\^o}t{\'e}}, \& {Lugaro}}]{Brinkman1}
{Brinkman}, H.~E., {Doherty}, C.~L., {Pols}, O.~R., {et~al.} 2019, \apj, 884,
  38

\bibitem[{{Chieffi} \& {Limongi}(2013)}]{LandC2013}
{Chieffi}, A., \& {Limongi}, M. 2013, \apj, 764, 21

\bibitem[{{C{\^o}t{\'e}} {et~al.}(2019{\natexlab{a}}){C{\^o}t{\'e}}, {Lugaro},
  {Reifarth}, {Pignatari}, {Vil{\'a}gos}, {Yag{\"u}e}, \&
  {Gibson}}]{Cote2019GCESLRs}
{C{\^o}t{\'e}}, B., {Lugaro}, M., {Reifarth}, R., {et~al.} 2019{\natexlab{a}},
  \apj, 878, 156

\bibitem[{{C{\^o}t{\'e}} {et~al.}(2019{\natexlab{b}}){C{\^o}t{\'e}},
  {Yag{\"u}e}, {Vil{\'a}gos}, \& {Lugaro}}]{Cote2019MonteCarlo}
{C{\^o}t{\'e}}, B., {Yag{\"u}e}, A., {Vil{\'a}gos}, B., \& {Lugaro}, M.
  2019{\natexlab{b}}, \apj, 887, 213

\bibitem[{{Cyburt} {et~al.}(2010){Cyburt}, {Amthor}, {Ferguson}, {Meisel},
  {Smith}, {Warren}, {Heger}, {Hoffman}, {Rauscher}, {Sakharuk}, {Schatz},
  {Thielemann}, \& {Wiescher}}]{CyburtJINA2010}
{Cyburt}, R.~H., {Amthor}, A.~M., {Ferguson}, R., {et~al.} 2010, \apjs, 189,
  240

\bibitem[{{de Smet} {et~al.}(2007){de Smet}, {Wagemans}, {Goeminne}, {Heyse},
  \& {van Gils}}]{deSmet2007}
{de Smet}, L., {Wagemans}, C., {Goeminne}, G., {Heyse}, J., \& {van Gils}, J.
  2007, \prc, 75, 034617

\bibitem[{{Diehl}(2013)}]{Diehl2013}
{Diehl}, R. 2013, Reports on Progress in Physics, 76, 026301

\bibitem[{{Dwarkadas} {et~al.}(2017){Dwarkadas}, {Dauphas}, {Meyer},
  {Boyajian}, \& {Bojazi}}]{Dwarkadas2017}
{Dwarkadas}, V.~V., {Dauphas}, N., {Meyer}, B., {Boyajian}, P., \& {Bojazi}, M.
  2017, \apj, 851, 147

\bibitem[{{Ekstr{\"o}m} {et~al.}(2012){Ekstr{\"o}m}, {Georgy}, {Eggenberger},
  {Meynet}, {Mowlavi}, {Wyttenbach}, {Granada}, {Decressin}, {Hirschi},
  {Frischknecht}, {Charbonnel}, \& {Maeder}}]{Ekstrom2012}
{Ekstr{\"o}m}, S., {Georgy}, C., {Eggenberger}, P., {et~al.} 2012, \aap, 537,
  A146

\bibitem[{{Farmer} {et~al.}(2016){Farmer}, {Fields}, {Petermann}, {Dessart},
  {Cantiello}, {Paxton}, \& {Timmes}}]{Farmer2016}
{Farmer}, R., {Fields}, C.~E., {Petermann}, I., {et~al.} 2016, \apjs, 227, 22

\bibitem[{{Gaidos} {et~al.}(2009){Gaidos}, {Krot}, {Williams}, \&
  {Raymond}}]{Gaidos2009}
{Gaidos}, E., {Krot}, A.~N., {Williams}, J.~P., \& {Raymond}, S.~N. 2009, \apj,
  696, 1854

\bibitem[{{G{\"o}tberg} {et~al.}(2018){G{\"o}tberg}, {de Mink}, {Groh},
  {Kupfer}, {Crowther}, {Zapartas}, \& {Renzo}}]{Gotberg2018Condition}
{G{\"o}tberg}, Y., {de Mink}, S.~E., {Groh}, J.~H., {et~al.} 2018, \aap, 615,
  A78

\bibitem[{{Gounelle} \& {Meibom}(2007)}]{GounelleMeibom2007}
{Gounelle}, M., \& {Meibom}, A. 2007, \apjl, 664, L123

\bibitem[{{Gounelle} \& {Meynet}(2012)}]{Gounelle2012}
{Gounelle}, M., \& {Meynet}, G. 2012, \aap, 545, A4

\bibitem[{{Hamann} {et~al.}(1995){Hamann}, {Koesterke}, \&
  {Wessolowski}}]{Hamann1995}
{Hamann}, W.-R., {Koesterke}, L., \& {Wessolowski}, U. 1995, \aap, 299, 151

\bibitem[{{Heger} {et~al.}(2000){Heger}, {Langer}, \& {Woosley}}]{Heger2000}
{Heger}, A., {Langer}, N., \& {Woosley}, S.~E. 2000, \apj, 528, 368

\bibitem[{{Heger} {et~al.}(2005){Heger}, {Woosley}, \& {Spruit}}]{Heger2005TS}
{Heger}, A., {Woosley}, S.~E., \& {Spruit}, H.~C. 2005, \apj, 626, 350

\bibitem[{{Imbriani} {et~al.}(2005){Imbriani}, {Costantini}, {Formicola},
  {Vomiero}, {Angulo}, {Bemmerer}, {Bonetti}, {Broggini}, {Confortola},
  {Corvisiero}, {Cruz}, {Descouvemont}, {F{\"u}l{\"o}p}, {Gervino},
  {Guglielmetti}, {Gustavino}, {Gy{\"u}rky}, {Jesus}, {Junker}, {Klug},
  {Lemut}, {Menegazzo}, {Prati}, {Roca}, {Rolfs}, {Romano}, {Rossi-Alvarez},
  {Sch{\"u}mann}, {Sch{\"u}rmann}, {Somorjai}, {Straniero}, {Strieder},
  {Terrasi}, \& {Trautvetter}}]{Imbriani2005}
{Imbriani}, G., {Costantini}, H., {Formicola}, A., {et~al.} 2005, European
  Physical Journal A, 25, 455

\bibitem[{{Jacobsen} {et~al.}(2008){Jacobsen}, {Yin}, {Moynier}, {Amelin},
  {Krot}, {Nagashima}, {Hutcheon}, \& {Palme}}]{Jacobsen2008}
{Jacobsen}, B., {Yin}, Q.-z., {Moynier}, F., {et~al.} 2008, Earth and Planetary
  Science Letters, 272, 353

\bibitem[{{Jones} {et~al.}(2019){Jones}, {M{\"o}ller}, {Fryer}, {Fontes},
  {Trappitsch}, {Even}, {Couture}, {Mumpower}, \& {Safi-Harb}}]{jones:19}
{Jones}, S.~W., {M{\"o}ller}, H., {Fryer}, C.~L., {et~al.} 2019, \mnras, 485,
  4287

\bibitem[{{J{\"o}nsson} {et~al.}(2014{\natexlab{a}}){J{\"o}nsson}, {Ryde},
  {Harper}, {Richter}, \& {Hinkle}}]{Joensson2014b}
{J{\"o}nsson}, H., {Ryde}, N., {Harper}, G.~M., {Richter}, M.~J., \& {Hinkle},
  K.~H. 2014{\natexlab{a}}, \apjl, 789, L41

\bibitem[{{J{\"o}nsson} {et~al.}(2017){J{\"o}nsson}, {Ryde}, {Spitoni},
  {Matteucci}, {Cunha}, {Smith}, {Hinkle}, \& {Schultheis}}]{Joensson2017}
{J{\"o}nsson}, H., {Ryde}, N., {Spitoni}, E., {et~al.} 2017, \apj, 835, 50

\bibitem[{{J{\"o}nsson} {et~al.}(2014{\natexlab{b}}){J{\"o}nsson}, {Ryde},
  {Harper}, {Cunha}, {Schultheis}, {Eriksson}, {Kobayashi}, {Smith}, \&
  {Zoccali}}]{Joensson2014a}
{J{\"o}nsson}, H., {Ryde}, N., {Harper}, G.~M., {et~al.} 2014{\natexlab{b}},
  \aap, 564, A122

\bibitem[{{Keszthelyi} {et~al.}(2020){Keszthelyi}, {Meynet}, {Shultz},
  {David-Uraz}, {ud-Doula}, {Townsend}, {Wade}, {Georgy}, {Petit}, \&
  {Owocki}}]{ZoltRotationalBoost}
{Keszthelyi}, Z., {Meynet}, G., {Shultz}, M.~E., {et~al.} 2020, \mnras, 493,
  518

\bibitem[{{Lamers} {et~al.}(1995){Lamers}, {Snow}, \&
  {Lindholm}}]{Lamer1995Alpha}
{Lamers}, H. J.~G.~L.~M., {Snow}, T.~P., \& {Lindholm}, D.~M. 1995, \apj, 455,
  269

\bibitem[{{Langer}(1998)}]{LangerBoost}
{Langer}, N. 1998, \aap, 329, 551

\bibitem[{{Lau} {et~al.}(2020){Lau}, {Eldridge}, {Hankins}, {Lamberts},
  {Sakon}, \& {Williams}}]{Lau2020Dust}
{Lau}, R.~M., {Eldridge}, J.~J., {Hankins}, M.~J., {et~al.} 2020, \apj, 898, 74

\bibitem[{{Limongi} \& {Chieffi}(2006)}]{LandC2006}
{Limongi}, M., \& {Chieffi}, A. 2006, \apj, 647, 483

\bibitem[{{Limongi} \& {Chieffi}(2018)}]{LandC2018}
---. 2018, \apjs, 237, 13

\bibitem[{{Liu}(2017)}]{Liu2017}
{Liu}, M.-C. 2017, \gca, 201, 123

\bibitem[{{Lodders}(2003)}]{Lodders2003}
{Lodders}, K. 2003, \apj, 591, 1220

\bibitem[{{Lugaro} {et~al.}(2018){Lugaro}, {Ott}, \& {Kereszturi}}]{Lugaro2018}
{Lugaro}, M., {Ott}, U., \& {Kereszturi}, {\'A}. 2018, Progress in Particle and
  Nuclear Physics, 102, 1

\bibitem[{{Luu} {et~al.}(2019){Luu}, {Hin}, {Coath}, \& {Elliott}}]{Luu2019}
{Luu}, T.-H., {Hin}, R.~C., {Coath}, C.~D., \& {Elliott}, T. 2019, Earth and
  Planetary Science Letters, 522, 166

\bibitem[{{Maeder} \& {Meynet}(2000{\natexlab{a}})}]{MandM2000RotationalBoost}
{Maeder}, A., \& {Meynet}, G. 2000{\natexlab{a}}, \aap, 361, 159

\bibitem[{{Maeder} \& {Meynet}(2000{\natexlab{b}})}]{MandM2000}
---. 2000{\natexlab{b}}, \araa, 38, 143

\bibitem[{{Maeder} \& {Meynet}(2012)}]{2012RvMP-MM}
---. 2012, Reviews of Modern Physics, 84, 25

\bibitem[{{Maeder} \& {Zahn}(1998)}]{1998maeder}
{Maeder}, A., \& {Zahn}, J.-P. 1998, \aap, 334, 1000

\bibitem[{{Makide} {et~al.}(2011){Makide}, {Nagashima}, {Krot}, {Huss},
  {Ciesla}, {Hellebrand}, {Gaidos}, \& {Yang}}]{Makide2011}
{Makide}, K., {Nagashima}, K., {Krot}, A.~N., {et~al.} 2011, \apjl, 733, L31

\bibitem[{{Meyer} \& {Clayton}(2000)}]{MeyerClayton2000}
{Meyer}, B.~S., \& {Clayton}, D.~D. 2000, \ssr, 92, 133

\bibitem[{{Meynet} \& {Arnould}(2000)}]{MeynetArnouldF192000}
{Meynet}, G., \& {Arnould}, M. 2000, \aap, 355, 176

\bibitem[{{Murray}(2011)}]{Murray2011}
{Murray}, N. 2011, \apj, 729, 133

\bibitem[{{Nesaraja} \& {McCutchan}(2016)}]{Cahalflife}
{Nesaraja}, C., \& {McCutchan}, E. 2016, Nuclear Data Sheets, 133, 120

\bibitem[{{Nica} {et~al.}(2012){Nica}, {Cameron}, \& {Singh}}]{Clhalflife}
{Nica}, N., {Cameron}, J., \& {Singh}, J. 2012, Nuclear Data Sheets, 113, 25

\bibitem[{{Nieuwenhuijzen} \& {de Jager}(1990)}]{NieuwenhuijzendeJager1990}
{Nieuwenhuijzen}, H., \& {de Jager}, C. 1990, \aap, 231, 134

\bibitem[{{Nugis} \& {Lamers}(2000)}]{NugisLamers2000}
{Nugis}, T., \& {Lamers}, H.~J.~G.~L.~M. 2000, \aap, 360, 227

\bibitem[{{O'Connor} \& {Ott}(2011)}]{CompactnessOconnor}
{O'Connor}, E., \& {Ott}, C.~D. 2011, \apj, 730, 70

\bibitem[{{Palacios} {et~al.}(2005){Palacios}, {Arnould}, \&
  {Meynet}}]{PalaciosF192005}
{Palacios}, A., {Arnould}, M., \& {Meynet}, G. 2005, \aap, 443, 243

\bibitem[{{Paxton} {et~al.}(2011){Paxton}, {Bildsten}, {Dotter}, {Herwig},
  {Lesaffre}, \& {Timmes}}]{MESA1}
{Paxton}, B., {Bildsten}, L., {Dotter}, A., {et~al.} 2011, \apjs, 192, 3

\bibitem[{{Paxton} {et~al.}(2013){Paxton}, {Cantiello}, {Arras}, {Bildsten},
  {Brown}, {Dotter}, {Mankovich}, {Montgomery}, {Stello}, {Timmes}, \&
  {Townsend}}]{MESA2}
{Paxton}, B., {Cantiello}, M., {Arras}, P., {et~al.} 2013, \apjs, 208, 4

\bibitem[{{Paxton} {et~al.}(2015){Paxton}, {Marchant}, {Schwab}, {Bauer},
  {Bildsten}, {Cantiello}, {Dessart}, {Farmer}, {Hu}, {Langer}, {Townsend},
  {Townsley}, \& {Tomes}}]{MESA3}
{Paxton}, B., {Marchant}, P., {Schwab}, J., {et~al.} 2015, \apjs, 220, 15

\bibitem[{{Paxton} {et~al.}(2018){Paxton}, {Schwab}, {Bauer}, {Bildsten},
  {Blinnikov}, {Duffell}, {Farmer}, {Goldberg}, {Marchant}, {Sorokina},
  {Thoul}, {Townsend}, \& {Timmes}}]{MESA4}
{Paxton}, B., {Schwab}, J., {Bauer}, E.~B., {et~al.} 2018, \apjs, 234, 34

\bibitem[{{Pignatari} {et~al.}(2016){Pignatari}, {Herwig}, {Hirschi},
  {Bennett}, {Rockefeller}, {Fryer}, {Timmes}, {Ritter}, {Heger}, {Jones},
  {Battino}, {Dotter}, {Trappitsch}, {Diehl}, {Frischknecht}, {Hungerford},
  {Magkotsios}, {Travaglio}, \& {Young}}]{Pignatari2016}
{Pignatari}, M., {Herwig}, F., {Hirschi}, R., {et~al.} 2016, VizieR Online Data
  Catalog, J/ApJS/225/24

\bibitem[{{Prantzos}(2012)}]{Prantzos2012CR}
{Prantzos}, N. 2012, \aap, 538, A80

\bibitem[{Rauscher(2008)}]{RauscherNonSmoker}
Rauscher, T. 2008, Online code NON-SMOKERWEB,version 5.0w and higher

\bibitem[{{Ritter} {et~al.}(2018){Ritter}, {Herwig}, {Jones}, {Pignatari},
  {Fryer}, \& {Hirschi}}]{Ritter2018}
{Ritter}, C., {Herwig}, F., {Jones}, S., {et~al.} 2018, \mnras, 480, 538

\bibitem[{{Rugel} {et~al.}(2009){Rugel}, {Faestermann}, {Knie}, {Korschinek},
  {Poutivtsev}, {Schumann}, {Kivel}, {G{\"u}nther-Leopold}, {Weinreich}, \&
  {Wohlmuther}}]{FeHalfLife}
{Rugel}, G., {Faestermann}, T., {Knie}, K., {et~al.} 2009, \prl, 103, 072502

\bibitem[{{Ryde} {et~al.}(2020){Ryde}, {J{\"o}nsson}, {Mace}, {Cunha},
  {Spitoni}, {Af{\textcommabelow s}ar}, {Jaffe}, {Forsberg}, {Kaplan},
  {Kidder}, {Lee}, {Oh}, {Smith}, {Sneden}, {Sokal}, {Strickland}, \&
  {Thorsbro}}]{RydeF192020}
{Ryde}, N., {J{\"o}nsson}, H., {Mace}, G., {et~al.} 2020, \apj, 893, 37

\bibitem[{{Schneider} {et~al.}(2021){Schneider}, {Podsiadlowski}, \&
  {M{\"u}ller}}]{Schneider2021Compactness}
{Schneider}, F.~R.~N., {Podsiadlowski}, P., \& {M{\"u}ller}, B. 2021, \aap,
  645, A5

\bibitem[{Sevior {et~al.}(1986)Sevior, Mitchell, Tingwell, \&
  Sargood}]{SEVIOR1986128}
Sevior, M., Mitchell, L., Tingwell, C., \& Sargood, D. 1986, Nuclear Physics A,
  454, 128

\bibitem[{{Stancliffe} {et~al.}(2005){Stancliffe}, {Lugaro}, {Ugalde}, {Tout},
  {G{\"o}rres}, \& {Wiescher}}]{Stancliffe2005}
{Stancliffe}, R.~J., {Lugaro}, M., {Ugalde}, C., {et~al.} 2005, \mnras, 360,
  375

\bibitem[{{Sukhbold} {et~al.}(2016){Sukhbold}, {Ertl}, {Woosley}, {Brown}, \&
  {Janka}}]{Sukhbold2016}
{Sukhbold}, T., {Ertl}, T., {Woosley}, S.~E., {Brown}, J.~M., \& {Janka}, H.-T.
  2016, \apj, 821, 38

\bibitem[{{Tang} \& {Dauphas}(2012)}]{TangDauphas2012}
{Tang}, H., \& {Dauphas}, N. 2012, Earth and Planetary Science Letters, 359,
  248

\bibitem[{{Tang} {et~al.}(2017){Tang}, {Liu}, {McKeegan}, {Tissot}, \&
  {Dauphas}}]{Tang2017}
{Tang}, H., {Liu}, M.-C., {McKeegan}, K.~D., {Tissot}, F. L.~H., \& {Dauphas},
  N. 2017, \gca, 207, 1

\bibitem[{{Trappitsch} {et~al.}(2018){Trappitsch}, {Boehnke}, {Stephan},
  {Telus}, {Savina}, {Pardo}, {Davis}, {Dauphas}, {Pellin}, \&
  {Huss}}]{Trappitsch18}
{Trappitsch}, R., {Boehnke}, P., {Stephan}, T., {et~al.} 2018, \apjl, 857, L15

\bibitem[{{Tur} {et~al.}(2010){Tur}, {Heger}, \& {Austin}}]{tur:10}
{Tur}, C., {Heger}, A., \& {Austin}, S.~M. 2010, \apj, 718, 357

\bibitem[{{Ugalde} {et~al.}(2008){Ugalde}, {Azuma}, {Couture}, {G{\"o}rres},
  {Lee}, {Stech}, {Strandberg}, {Tan}, \& {Wiescher}}]{Ugalde2008}
{Ugalde}, C., {Azuma}, R.~E., {Couture}, A., {et~al.} 2008, \prc, 77, 035801

\bibitem[{{Vink} \& {de Koter}(2005)}]{VinkdeKoter2005}
{Vink}, J.~S., \& {de Koter}, A. 2005, \aap, 442, 587

\bibitem[{{Vink} {et~al.}(2000){Vink}, {de Koter}, \& {Lamers}}]{Vink2000}
{Vink}, J.~S., {de Koter}, A., \& {Lamers}, H.~J.~G.~L.~M. 2000, \aap, 362, 295

\bibitem[{{Vink} {et~al.}(2001){Vink}, {de Koter}, \& {Lamers}}]{Vink2001}
---. 2001, \aap, 369, 574

\bibitem[{{Wang} {et~al.}(2020){Wang}, {Siegert}, {Dai}, {Diehl}, {Greiner},
  {Heger}, {Krause}, {Lang}, {Pleintinger}, \& {Zhang}}]{Wang2020Iron60}
{Wang}, W., {Siegert}, T., {Dai}, Z.~G., {et~al.} 2020, \apj, 889, 169

\bibitem[{{Wasserburg} {et~al.}(2006){Wasserburg}, {Busso}, {Gallino}, \&
  {Nollett}}]{Wasserburg2006}
{Wasserburg}, G.~J., {Busso}, M., {Gallino}, R., \& {Nollett}, K.~M. 2006,
  \nphysa, 777, 5

\bibitem[{{Young}(2014)}]{Young2014}
{Young}, E.~D. 2014, Earth and Planetary Science Letters, 392, 16

\end{thebibliography}
\end{document}